\documentclass[12pt,a4paper]{article}

\usepackage{amsmath,amssymb,bbm} 
\usepackage{accents}
\usepackage{color}
\usepackage[bookmarks=true,hyperfigures=true]{hyperref}
\usepackage{graphicx}
\usepackage[nottoc,notlot,notlof]{tocbibind}
\usepackage[nosort]{cite}
\usepackage[bulletsep]{collref}
\usepackage{tensor}

\usepackage{color}

\usepackage[a4paper,text={173mm,216mm},centering]{geometry}

\allowdisplaybreaks[3]

\usepackage[font=small,labelfont=bf,width=0.85\textwidth]{caption}

\let\oldbfseries=\bfseries
\let\oldmdseries=\mdseries
\let\oldnormalfont=\normalfont
\renewcommand{\bfseries}{\oldbfseries\boldmath}
\renewcommand{\mdseries}{\oldmdseries\unboldmath}
\renewcommand{\normalfont}{\oldnormalfont\unboldmath}

\makeatletter
\newlength{\apb@width}
\newcommand{\autoparbox}[2][c]{\settowidth{\apb@width}{#2}\parbox[#1]{\apb@width}{#2}}

\makeatother

\newcommand{\nn}{\nonumber}

\newcommand{\remark}[2][.]{{\color{red}\renewcommand{\bfdefault}{b}\rmfamily\if.#1\else\textbf{#1:} \fi#2}}

\newcommand{\be}{\begin{equation}}
\newcommand{\ee}{\end{equation}}
\newcommand{\ba}{\begin{eqnarray}}
\newcommand{\ea}{\end{eqnarray}}

\ifx\genfrac\sdflkaj\else\fi

\renewcommand{\a}{\alpha}
\newcommand{\da}{{\dot{\alpha}}}

\newcommand{\la}{\lambda}
\newcommand{\tla}{\tilde\lambda}

\newcommand{\ft}[2]{{\textstyle\frac{#1}{#2}}}

\newcommand{\cA}{\mathcal{A}}
\newcommand{\cM}{\mathcal{M}}
\newcommand{\cO}{\mathcal{O}}

\def\l<{\langle}\def\r>{\rangle}

\def\bra(#1,#2){[#1\,#2]}
\def\ket(#1,#2){\langle #1\,#2\rangle}

\makeatletter
\newcommand{\namedref}[2]{\hyperref[#2]{#1~\ref*{#2}}}
\newcommand{\secref}{\@ifstar{\namedref{Section}}{\namedref{sec.}}}
\newcommand{\subsecref}{\@ifstar{\namedref{Subsection}}{\namedref{subsec.}}}
\newcommand{\appref}{\@ifstar{\namedref{Appendix}}{\namedref{app.}}}
\newcommand{\tabref}{\@ifstar{\namedref{Table}}{\namedref{tab.}}}
\newcommand{\figref}{\@ifstar{\namedref{Figure}}{\namedref{fig.}}}
\makeatother

\newif\ifmrnote 
\mrnotetrue
 
\usepackage[active]{srcltx}

\newif\ifjbnote 
\jbnotetrue
 
\def\[{\begin{equation}}
\def\]{\end{equation}}
\def\<{\begin{eqnarray}}
\def\>{\end{eqnarray}}

\newcommand{\biggbrk}[1]{\biggl(#1\biggr)}

\newcommand{\biggsbrk}[1]{\biggl[#1\biggr]}

\newcommand{\vev}[1]{\langle#1\rangle}
\newcommand{\bev}[1]{ [#1]}

\def\tlaDtla(#1,#2){\tla_{#1}^{\dot{\alpha}} \frac{\partial}{\partial{\tla}^{\dot {\alpha}}_{#2}}}
\def\laDla(#1,#2){\la_{#1}^{ \alpha} \frac{\partial}{\partial{\la}^{\alpha}_{#2}}}

\def\beqa{\begin{eqnarray}}
\def\eeqa{\end{eqnarray}}
\def\beq{\begin{equation}}
\def\eeq{\end{equation}}

\begin{document}
\thispagestyle{empty}


\begin{flushright}
{\tt HU-EP-15/17},
{\tt HU-MATH-2015-03}\\
{\tt TCDMATH 15-03},
{\tt QMUL-PH-15-08}
\end{flushright}

\begin{center}
{\LARGE\bfseries Double-Soft Limits of 
Gluons and Gravitons
\par}%
\vspace{10mm}

\begingroup\scshape\large 
Thomas Klose${}^{1}$, Tristan McLoughlin${}^{2}$,  Dhritiman Nandan${}^{1,5}$, \\[0.3cm] Jan Plefka${}^{1}$ and
Gabriele Travaglini${}^{1,3,4}$
\endgroup
\vspace{5mm}

\textit{${}^{1}$ Institut f\"ur Physik und IRIS Adlershof, Humboldt-Universit\"at zu Berlin \phantom{$^\S$}\\
  Zum Gro{\ss}en Windkanal 6, 12489 Berlin, Germany} \\[0.1cm]
\texttt{\small \{thklose,dhritiman,plefka\}@physik.hu-berlin.de\phantom{\ldots}} \vspace{4mm}

\textit{${}^{2}$ School of Mathematics, Trinity College Dublin\\ College Green, Dublin 2, Ireland}\\[0.1cm]
  \texttt{\small tristan@maths.tcd.ie\phantom{\ldots}} \\ \vspace{4mm}

\textit{${}^{3}$ Centre for Research in String Theory\\ School of Physics and Astronomy, Queen Mary University of London\\
Mile End Road, London E1 4NS, United Kingdom}\\[0.1cm]
  \texttt{\small g.travaglini@qmul.ac.uk \phantom{\ldots}} \\ \vspace{4mm}

\textit{${}^{4}$ Dipartimento di Fisica, Universit\`{a} di Roma ``Tor Vergata" \\
Via della Ricerca Scientifica, 1 00133 Roma, Italy}\\[0.1cm]
 \vspace{4mm}

\textit{${}^{5}$ Institut f\"ur Mathematik und IRIS Adlershof, Humboldt-Universit\"at zu Berlin \phantom{$^\S$}\\
  Zum Gro{\ss}en Windkanal 6, 12489 Berlin, Germany} \\[0.1cm]
  \vspace{4mm}

\textbf{Abstract}\vspace{5mm}\par
\begin{minipage}{14.7cm}

The double-soft limit of gluon and graviton amplitudes is studied in four dimensions
at tree level. In general this limit is ambiguous and we introduce two natural ways
of taking it: A consecutive double-soft limit where one particle is
taken soft before the other and a simultaneous limit where both particles are taken soft 
uniformly. All limits yield universal factorisation formulae which we establish by BCFW
recursion relations down to the subleading order in the soft momentum expansion. These
formulae generalise the recently discussed subleading single-soft theorems. While both types
of limits yield identical results at the leading order, differences appear at the subleading order.
Finally, we  discuss double-scalar emission in $\mathcal{N}\!=\!4$ super Yang-Mills theory. 
These results should be of use in  establishing the algebraic structure of potential hidden
symmetries in the quantum gravity and Yang-Mills S-matrix.

\end{minipage}\par
\end{center}
\newpage


\setcounter{tocdepth}{4}
\hrule height 0.75pt
\tableofcontents
\vspace{0.8cm}
\hrule height 0.75pt
\vspace{1cm}

\setcounter{tocdepth}{2}



\section{Introduction and Conclusions}

The infrared behaviour of gluon and graviton amplitudes displays
a universal factorisation into a soft and a hard contribution which makes it an 
interesting topic of study. As
was already noticed in the early days of quantum field theory 
\cite{Low:1958sn,Weinberg:1964ew}, the emission of a single soft gluon 
or graviton yields a singular soft function linearly divergent in the soft momentum. There is also universal behaviour at the subleading
order in a soft momentum expansion both for gluons and photons \cite{Low:1958sn,Burnett:1967km,Casali:2014xpa}
and, as was discovered only recently, for gravitons \cite{Cachazo:2014fwa}. 
The authors of \cite{Cachazo:2014fwa} moreover related the
subleading soft graviton functions to a conjectured hidden symmetry of the quantum gravity
S-matrix \cite{Kapec:2014opa,Strominger:2014pwa} which has the form of an extended BMS${}_{4}$ algebra 
\cite{Bondi:1962px,Sachs:1962wk,Barnich:2010eb} known from 
classical gravitational waves. Similar claims that the Yang-Mills S-matrix enjoys a hidden 
two-dimensional Kac-Moody type symmetry were made recently \cite{He:2015zea}. In this picture the scattering amplitudes in
four-dimensional quantum field theory are related to correlation functions 
of a two-dimensional quantum theory living on the sphere at null infinity. This fascinating
proposal merits further study. 

The subleading soft gluon and graviton theorems were proven using modern on-shell 
techniques for scattering amplitudes\footnote{See e.g.~\cite{Elvang:2013cua,Henn:2014yza} for a textbook treatment.}. 
They hold in general dimensions 
\cite{Schwab:2014xua,Afkhami-Jeddi:2014fia,Zlotnikov:2014sva,Kalousios:2014uva} and their form
is strongly constrained by gauge and Poincar\'e symmetry \cite{Broedel:2014fsa,Bern:2014vva}. These results are
so far restricted to tree-level. The important loop-level validity
and deformations of the theorem were studied in
\cite{Bern:2014oka,He:2014bga,Cachazo:2014dia,Bianchi:2014gla,Broedel:2014bza}. 
An ambitwistor string model was proposed in \cite{Geyer:2014lca,Adamo:2014yya} which yields the graviton and gluon
tree-level S-matrix in the form of their CHY representation 
\cite{Cachazo:2013hca,Cachazo:2013iea}. In this language the
soft theorems have an intriguing two-dimensional origin in terms of corresponding limits of the vertex operators 
on the ambitwistor string world-sheet \cite{Lipstein:2015rxa}.

Technically the soft theorems are conveniently expressed as an expansion in a small soft scaling parameter
$\delta$ multiplying the momentum of the soft particle $p^{\mu}=\delta\, q^{\mu}$
with $q^{2}=0$. Taking the soft limit of a gluon in a colour-ordered $(n+1)$-point amplitude $A_{n+1}$ yields the soft theorem at tree-level
\be
\lim_{\delta\to 0} \cA_{n+1} = \Bigl (\frac 1 \delta \, S^{(0)}_{\text{YM}}(q)+ S^{(1)}_{\text{YM}}(q)
\Bigr)\, \cA_{n}
+ \cO(\delta) \, ,
\ee
where $\cA_{n}=\delta^{(4)}(\sum_{i=1}^{n}p_{n})\, A_{n}$ denotes the full amplitude including the 
momentum preserving delta-function. The soft functions $S^{(n)}_{\text{YM}}(q)$ are universal,
in fact $S^{(1)}_{\text{YM}}(q)$ has the form of a differential operator in momenta and 
polarisations acting on the
the amplitude $\cA_{n}$.
For soft gravitons the universality even extends down to the sub-subleading
order 
\be
\lim_{\delta\to 0} \cM_{n+1} = \Bigl (\frac 1 \delta \, S^{(0)}_{\text{grav}}(q)+ S^{(1)}_{\text{grav}}(q)
+ \delta\,  S^{(2)}_{\text{grav}}(q)\Bigr)\, \cM_{n}
+ \cO(\delta^{2}) \, .
\ee
Now $S^{(1)}_{\text{grav}}$ is a first-order and $S^{(2)}_{\text{grav}}$ a second-order
differential operator in the hard momenta and polarisations (or equivalently in spinor helicity variables).
The leading soft function $S^{(0)}_{\text{grav}}$ has been associated 
\cite{Kapec:2014opa,Strominger:2014pwa} to the Ward identity of the super-translation, while the subleading soft function $S^{(1)}_{\text{grav}}$ 
to that of the Virasoro (or super-rotation) generators of the extended BMS${}_{4}$ symmetry algebra.
However, this subleading connection is still not entirely established.

The soft behaviour of the S-matrix is in general connected to its symmetries. 
Hence exploring the soft behaviour is a means to uncover hidden symmetries in quantum
field theories.
This is particularly
transparent in the soft behaviour of Goldstone bosons of a spontaneously broken symmetry. 
In this situation the soft limit of a single scalar in the theory leads to a vanishing amplitude known as Adler's zero
\cite{Adler:1964um}. The emergence of a hidden  symmetry algebra from the soft behaviour of 
amplitudes has been beautifully demonstrated in \cite{ArkaniHamed:2008gz}: Taking the \emph{double} soft
 limit for two scalars reveals the algebraic structure and yields a non-vanishing result 
of the form
\be
\lim_{\delta\to 0} \cA_{n+2}(\phi^{i}(\delta q_{1}), \phi^{j}(\delta q_{2}), 3,\ldots n+2) = 
\sum_{a=3}^{n+2} \frac{p_{a}\cdot(q_{1}-q_{2})}{p_{a}\cdot(q_{1}+q_{2})}\, f^{ijK}T_{K}\cA_{n}(3,\ldots n+2) + \cO(\delta)
\ee
where $T_{K}$ is the generator of the invariant subgroup with $[T^{i},T^{j}]=f^{ijK}T_{K}$
in a suitable representation for acting on amplitudes.
Using this method the authors of \cite{ArkaniHamed:2008gz} demonstrated that the double-soft
limit of two scalars in $\mathcal{N}=8$ supergravity gives rise to the structure constants
of the hidden $E_{7(7)}$ symmetry algebra acting non-linearly on the scalars. Methods for extracting double-soft limits of scalars in $4\leq\mathcal{N} <8$ supergravity as well as $\mathcal{N}=16$ supergravity in three dimensions were presented in \cite{Chen:2014cuc}.
Single soft scalar limits were also studied as a classification tool for effective field
theories in \cite{Cheung:2014dqa}.
Recently, the double-soft limits of spin $1/2$ particles were studied in a series of theories
and related universal double-soft behaviour could be established \cite{Chen:2014xoa}. 
Of course, for fermions the single-soft limit vanishes by statistics. Double-soft scalar and photon limits
were studied very recently for several classes of four-dimensional theories containing scalar particles 
in \cite{Cachazo:2015ksa} using the CHY representation \cite{Cachazo:2013hca,Cachazo:2013iea}. Interesting universal double-soft theorems were established.

In summary these results indicate that (i) double-soft limits of massless particles exhibit
universal behaviour going beyond the single-soft theorems, and (ii) that the double-soft limits
have the potential to exhibit the algebraic structure of underlying hidden symmetries of 
the S-matrix.
These insights and results set the stage for the present analysis where we lift the universal 
double-soft theorems of massless spin 0 and spin $1/2$ particles to the spin 1 and 2 cases.
The central difference now lies in the non-vanishing single-soft limits reviewed above. This
entails an ambiguity in the way one takes a double-soft limit of two gluons or gravitons with
momenta $\delta_{1} q_{1}$ and $\delta_{2}q_{2}$:
\begin{itemize}
\item
One can take a consecutive soft limit in which one first takes $\delta_{2}$ to zero and thereafter
$\delta_{1}$. 
\be
{\tt CSL}(1, 2){\cA}_n(3,\dots, n+2) = \lim_{\delta_1 \to 0} \lim_{\delta_2 \to 0} \cA_{n+2}(\delta_1 q_1, \delta_2 q_2, 3, \dots, n+2) \,.
\ee
The ambiguity of this limit is then reflected in a non vanishing anti-symmetrised
version of this consecutive limit
\be
{\tt aCSL}(1, 2){\cA}_n(3,\dots, n+2) = \ft 12 [ \lim_{\delta_1 \to 0} ,\lim_{\delta_2 \to 0} ]\, \cA_{n+2}(\delta_1 q_1, \delta_2 q_2, 3, \dots, n+2) \, .
\ee
In fact we shall see that for gluons or gravitons of the same
helicity the anti-symmetrised consecutive limit always vanishes at leading order. For the case of different
helicities of the two soft particles,  the anti-symmetrised consecutive limit is non-zero.
Such an anti-symmetrised consecutive limit for the case of identical helicity photons
and gravitons was recently studied in \cite{Lipstein:2015rxa}.
\item
Alternatively one can take a simultaneous soft limit in which one sets $\delta_{1}=\delta_{2}=\delta$
and sends both momenta simultaneously to zero
\be
{\tt DSL}(1, 2){\cA}_n(3,\dots, n+2) = \lim_{\delta\to 0}  \cA_{n+2}(\delta q_1, \delta q_2, 3, \dots, n+2) \,.
\ee
It is this limit which naturally arises in the scalar scenarios where a single soft limit
vanishes due to Adler's zero,  and thus also the consecutive double-soft limit.
\end{itemize}
Both double-soft functions have a leading quadratic divergence in the soft limit. In order
to obtain a uniform description we set $\delta_{1}=\delta_{2}=\delta$ also for the consecutive
limit \emph{after} having taken the limits. It is then natural to define the subleading
double-soft functions via the series
\be
{\tt CSL}(1,2) = \sum_{i} \delta^{i-2} {\tt CSL}^{(i)}(1,2)~ \quad \text{and}\quad 
{\tt DSL}(1,2) = \sum_{i} \delta^{i-2} {\tt DSL}^{(i)}(1,2)~ \, .
\ee
Universality extends down at least to the subleading order.

It is interesting to compare the two soft-functions. As we shall show at leading order in the case of identical helicities
of particles 1 and 2 they agree
\be
{\tt CSL}^{(0)}(1^{h},2^{h}) = {\tt DSL}^{(0)}(1^{h},2^{h}) \, .
\ee
both for gravity and Yang-Mills. At the subleading order still for the same helicities
the two continue to agree in the gravity case but differ in the colour-ordered Yang-Mills
case
\be
{\tt CSL}_{\text{gravity}}^{(1)}(1^{h},2^{h}) = {\tt DSL}_{\text{gravity}}^{(1)}(1^{h},2^{h}) 
\,  \quad \text{but}\qquad
{\tt CSL}_{\text{YM}}^{(1)}(1^{h},2^{h}) \neq {\tt DSL}_{\text{YM}}^{(1)}(1^{h},2^{h})
\, .
\ee
If the two soft particles carry opposite helicities the situation is different. While the
leading contributions continue to agree for gravity they now disagree at the leading level
also for Yang-Mills
\be
{\tt CSL}_{\text{gravity}}^{(0)}(1^{h},2^{\bar h}) = {\tt DSL}^{(0)}_{\text{gravity}}(1^{h},2^{\bar h})\, 
\quad \text{but} \quad
{\tt CSL}_{\text{YM}}^{(0)}(1^{h},2^{\bar h}) \neq {\tt DSL}^{(0)}_{\text{YM}}(1^{h},2^{\bar h}) \, .
\ee
At the subleading order both gravity and Yang-Mills disagree 
\be
{\tt CSL}_{\text{gravity}}^{(1)}(1^{h},2^{\bar h}) \neq {\tt DSL}_{\text{gravity}}^{(1)}(1^{h},2^{\bar h}) 
\,  \quad \text{and}\qquad
{\tt CSL}_{\text{YM}}^{(1)}(1^{h},2^{\bar h}) \neq {\tt DSL}_{\text{YM}}^{(1)}(1^{h},2^{\bar h})
\, .
\ee
These results should be of use for 
establishing the algebraic structure of potential hidden symmetries in the quantum gravity and Yang-Mills S-matrix. This, however, is left for future work.

As a final application of our work, we use  supersymmetric recursion relations \cite{ArkaniHamed:2008gz,Brandhuber:2008pf} in $\mathcal{N}=4$ super Yang-Mills to address double-soft limits. This set-up can be used to re-derive the double-soft limits of gluons obtained from the non-supersymmetric recursion relations, but also to study double-soft scalar emission. The interesting observation here is that while a single-soft scalar limit in $\mathcal{N}=4$ super Yang-Mills is finite, and hence non-universal, double-soft scalar emissions gives rise to a divergence, and we compute the corresponding double-soft scalar function. 

The paper is organised as follows. In the next section we first review single-soft limits of gluons and gravitons, and we then apply these results to study consecutive double-soft limits of the same particles. Section 3 and 4 contain the main results of this paper, namely the analysis of simultaneous double-soft limits of gluons and gravitons. Finally, we discuss double-soft scalar emission in Section 4. Two appendices with technical details of some of our calculations complete the paper.

\bigskip
{\bf Note added:} After finishing this work, we were made aware in recent email 
correspondence with  Anastasia Volovich and Congkao Wen of a work of Volovich, Wen and Zlotnikov \cite{Volovich:2015yoa} which has some overlap with our paper.

\section{Single and consecutive double-soft limits}

We start from an amplitude of $n\!+\!1$ particles with momenta $p_1$ to $p_{n+1}$ and take the momentum of the first particle to be soft by setting $p_1 = \delta_1 q_1$ and expanding the amplitude in powers of $\delta_1$. In terms of spinor variables, we define the soft limit by $\lambda_{p_1} = \sqrt{\delta_1} \lambda_{q_1}$ and $\tilde{\lambda}_{p_1} = \sqrt{\delta_1} \tilde{\lambda}_{q_1}$. \\
In order to keep the notation compact, we will use $\lambda_{q_1} \equiv \lambda_1 \equiv |1\rangle$ and $\tilde{\lambda}_{q_1} \equiv \tilde{\lambda}_1 \equiv |1]$ for the soft particle and $\lambda_{p_a} \equiv \lambda_a \equiv |a\rangle$ and $\tilde{\lambda}_{p_a} \equiv \tilde{\lambda}_a \equiv |a]$ for the hard ones $a=2,\ldots,n+1$.

\subsection{Single-soft limits}

\paragraph{Yang-Mills.}

The single-soft limit, including the subleading term, for color-ordered Yang-Mills amplitudes is given by 
\cite{Low:1958sn,Burnett:1967km,Casali:2014xpa}
\<
A_{n+1}(1^{h_1}, 2, \dots, n+1) = \biggsbrk{\frac{1}{\delta_1} S^{(0)}(n+1, 1^{h_1}, 2) + S^{(1)}(n+1, 1^{h_1}, 2) + \ldots}{A}_n(2, \dots, n+1) \, , 
\>
with 
\<
S^{(0)}(n+1, 1^+, 2) = \frac{\langle n\!+\!1 \, 2\rangle }{\langle n\!+\!1\, 1 \rangle \langle 1 2 \rangle}~, ~~~
S^{(1)}(n+1, 1^+, 2) = \frac{1}{\langle 1 2\rangle}{\tilde \lambda}_1^{\dot \alpha} \frac{\partial}{\partial{\tilde \lambda}^{\dot \alpha}_{2}} + \frac{1}{\langle n\!+\!1\, 1 \rangle }{\tilde \lambda}_1^{\dot \alpha} \frac{\partial}{\partial{\tilde \lambda}^{\dot \alpha}_{n\!+\!1}}
\>
for a positive-helicity gluon. For a negative-helicity gluon the soft factors are given by conjugation of the spinor variables, $\lambda_i \leftrightarrow {\tilde \lambda}_i$.

\paragraph{Gravity.}
For the gravitational case we have \cite{Weinberg:1964ew,Cachazo:2014fwa}
\<
{\cal M}_{n+1}(1^{h_1}, 2, \dots, n+1) = \biggsbrk{ \frac{1}{\delta_1} S^{(0)}(1^{h_1})+S^{(1)}(1^{h_1}) + \delta_1  S^{(2)}(1^{h_1}) + \ldots}{\cal M}_n(2, \dots, n+1)\, , 
\>
where for a positive-helicity graviton  
\<
\label{singlesoftgrp}
S^{(0)}(1^{+}) = \sum_{a=2}^{n+1}\frac{[1 a]}{\langle 1 a\rangle}\frac{\langle x a\rangle}{\langle x 1 \rangle} \frac{\langle y a\rangle }{\langle y 1 \rangle}~,~~~
S^{(1)}(1^{+}) = \frac{1}{2}\sum_{a=2}^{n+1}\frac{[1 a]}{\langle 1 a\rangle}\left(\frac{\langle x a\rangle}{\langle x 1 \rangle} + \frac{\langle y a\rangle }{\langle y 1 \rangle}\right){\tilde \lambda}^{\dot \alpha}_{1} \frac{\partial}{\partial{\tilde \lambda}_a^{\dot \alpha}}\, .
\>
The spinors $\lambda_{x}$ and $\lambda_{y}$ are arbitrary reference spinors. The sub-subleading term is given by 
\<
S^{(2)}(1^{+}) = \frac{1}{2} \sum_{a=2}^{n+1} \frac{[1 a]}{\langle 1 a\rangle} {\tilde \lambda}_1^{\dot \alpha} {\tilde \lambda}_1^{\dot \beta} \frac{\partial^2}{\partial {\tilde \lambda}_a^{\dot \alpha} \partial {\tilde \lambda}_a^{\dot \beta}}~.
\>
As for the gluonic case, the opposite helicity factors are found by conjugation.

\subsection{Consecutive double-soft limits}
\label{consec-sec}

In all double-soft limits, we start from an amplitude of $n\!+\!2$ particles and set the momenta of the first and the second particle to $p_1 = \delta_1 q_1$ and $p_2 = \delta_2 q_2$ respectively. In terms of spinor variables, we distribute the $\delta$'s symmetrically as above: $\{\sqrt{\delta_1} \lambda_{q_1}, \sqrt{\delta_1} {\tilde \lambda}_{q_1}\}$ and $\{\sqrt{\delta_2} \lambda_{q_2}, \sqrt{\delta_2} {\tilde \lambda}_{q_2}\}$. \\
By expanding the amplitude in $\delta_1$ and $\delta_2$, we obtain various double-soft limits. In the consecutive soft limit --- in contradistinction to the simultaneous double-soft limit to be discussed in the next section --- we first expand in $\delta_2$ while keeping $\delta_1$ fixed, and then expand each term of the series in $\delta_1$. The result can be calculated straightforwardly from repeated use of the above single-soft limits.

\paragraph{Yang-Mills.}

As above, we first consider the case of gluons. Let us define the ``consecutive soft limit factor''
${\tt CSL}(n+2, 1^{h_1}, 2^{h_2}, 3)$ by
\<
{\tt CSL}(n+2, 1^{h_1}, 2^{h_2},3 ){A}_n(3,\dots, n+2) &\equiv& \lim_{\delta_1 \to 0} \lim_{\delta_2 \to 0} A_{n+2}(\delta_1 q^{h_1}_1, \delta_2 q^{h_2}_2, 3, \dots, n+2) \nn\\
& &\kern-220pt = \biggsbrk{ \frac{1}{\delta_2} S^{(0)}(1, 2^{h_2}, 3)+S^{(1)}(1, 2^{h_2}, 3)} 
                 \biggsbrk{ \frac{1}{\delta_1} S^{(0)}(n+2, 1^{h_1}, 3)+S^{(1)}(n+2, 1^{h_1}, 3)} {A}_n(3,\dots, n+2)~.\nn
\>
We can also define symmetrised and antisymmetrised versions of the consecutive limits 
\<
{\tt sCSL}(n+2, 1^{h_1}, 2^{h_2}, 3){A}_n(3,\dots, n+2)&\equiv&\tfrac{1}{2}\{\lim_{\delta_1 \to 0}, \lim_{\delta_2 \to 0}\} {A}_{n+2}(\delta_1 q^{h_1}_1, \delta_2 q^{h_2}_2, 3 \dots, n+2)\, , \nn\\
{\tt aCSL}(n+2, 1^{h_1}, 2^{h_2}, 3){A}_n(3,\dots, n+2)&\equiv&\tfrac{1}{2}[\lim_{\delta_1 \to 0}, \lim_{\delta_2 \to 0}] {A}_{n+2}(\delta_1 q^{h_1}_1, \delta_2 q^{h_2}_2, 3 \dots, n+2)
~.
\>
As it will be of interest later, let us consider specific helicities:
\<
 {\tt CSL}(n+2, 1^{+}, 2^{+}, 3) &=& \frac{1}{\delta_1\delta_2}\frac{\langle n\!+\!2\, 3\rangle}{\langle n\!+\!2\, 1\rangle  \langle 1 2\rangle \langle 2 3\rangle}+{\cal O}(\delta_2^0/\delta_1, \delta_1^0/\delta_2)~,\nn\\
  {\tt CSL}(n+2, 1^{+}, 2^{-}, 3) &=& \frac{1}{\delta_1\delta_2}\frac{\langle n\!+\!2\, 3\rangle }{\langle  n\!+\!2\, 1\rangle [1 2] [2 3]}\frac{[1 3]}{\langle 1 3\rangle}+{\cal O}(\delta_2^0/\delta_1, \delta_1^0/\delta_2)~.
 \>
If we take the reverse consecutive limit, i.e.\ expand first in $\delta_1$ and then in $\delta_2$, the leading term in ${\tt CSL}(1^{+}, 2^{+})$ is unchanged; hence the symmetric combination is the same as either ordering while the antisymmetric combination vanishes. 
 
It is in fact useful to consider subleading terms; for simplicity, after expanding, we will set $\delta_1=\delta_2=\delta$ and define 
\<
{\tt CSL}(n+2,1^{h_1}, 2^{h_2},3) = \sum_{i} \delta^{i-2} {\tt CSL}^{(i)}(n+2, 1^{h_1}, 2^{h_2}, 3)~, 
\>
and similarly for ${\tt s/aCSL}$. The first subleading term is given by
\<
{\tt CSL}^{(1)}(n+2, 1^{+}, 2^{+}, 3) = S^{(0)}(1, 2^+, 3) S^{(1)}(n+2, 1^+, 3)+S^{(1)}(1, 2^+, 3) S^{(0)}(n+2, 1^+, 3)~.
\>
As $S^{(1)}$ involves derivatives there will in principle be ``contact'' terms when they act on the other soft factor, however as the derivatives are only with respect to the $\tilde \lambda$'s and $S^{(0)}$ depends only on the $\lambda$'s they are trivially zero%
\footnote{ 
It is perhaps worthwhile to note that this is only valid for generic external momenta as we neglect holomorphic anomaly terms that can arise when external legs are collinear with soft legs.
}.

A short calculation yields the symmetric and antisymmetric combination of the consecutive soft factor at the next order
 \<
   {\tt s/aCSL}^{(1)}(n+2, 1^{+}, 2^{+}, 3)&=&
   +\frac{1}{2}\biggbrk{\frac{\langle n\!+\!2\,  3\rangle \langle 1 2\rangle \pm \langle n\!+\!2\, 2\rangle \langle 1 3\rangle}{\langle 2 3\rangle \langle n\!+\!2\,  1\rangle \langle 1 2\rangle \langle 1 3\rangle }}
 {\tilde   \lambda}_{2}^{\dot \alpha}\frac{\partial}{\partial {\tilde \lambda}_3^{\dot \alpha}}\nn\\
   & &+\frac{1}{2}\biggbrk{\frac{\langle n\!+\!2\,  2\rangle \langle 1 3\rangle \pm \langle n\!+\!2\,  3\rangle \langle 1 2\rangle}{\langle 2 3\rangle \langle n\!+\!2\,  1\rangle \langle 1 2\rangle \langle n\!+\!2\,  2\rangle}}
   {\tilde \lambda}_{1}^{\dot \alpha}\frac{
   \partial}{\partial {\tilde \lambda}_{n+2}^{\dot \alpha}}\nn\\
   & &+\frac{1}{2}\frac{{\tilde \lambda}_{1}^{\dot \alpha}}{\langle 1 2\rangle \langle 2 3\rangle }
   \frac{\partial}{\partial {\tilde \lambda}_3^{\dot \alpha}}\pm \frac{1}{2}\frac{{\tilde \lambda}^{\dot \alpha}_{2}}{\langle n\!+\!2\,  1\rangle\langle 1 2 \rangle} \frac{
   \partial}{\partial {\tilde \lambda}_{n+2}^{\dot \alpha}}\, , 
 \>
where the upper sign corresponds to the symmetric case and the lower sign to the antisymmetric case. In the antisymmetric case, the expression can be simplified further, 
\<
  {\tt aCSL}^{(1)}(n+2, 1^{+}, 2^{+}, 3)& &
  \nn\\
 & & \kern-80pt =\frac{1}{2\langle 1 2\rangle}\bigg[
  \bigg(\frac{{\tilde \lambda}^{\dot \alpha}_{1}}{ \langle 2 3\rangle}-\frac{{\tilde \lambda}^{\dot \alpha}_{2}}{ \langle 1 3\rangle}\bigg)
 \frac{\partial}{\partial {\tilde \lambda}_3^{\dot \alpha}}
 -
  \bigg(\frac{{\tilde \lambda}^{\dot \alpha}_{1}}{ \langle 2\, n+2\rangle}-\frac{{\tilde \lambda}^{\dot \alpha}_{2}}{ \langle 1\, n+2\rangle}\Big)
 \frac{\partial}{\partial {\tilde \lambda}_{n+2}^{\dot \alpha}}
 \bigg]~.
\>
Turning to the case of mixed helicity, the leading term for the reversed limit is already different  and so we find 
 \<
 \label{eq:symm_mix_hel}
  {\tt s/aCSL}^{(0)}(n+2, 1^{+}, 2^{-}, 3) &=& \frac{1}{2}\frac{1}{\langle n\!+\!2 \,1\rangle [2 3]}\left(\frac{\langle n\!+\!2 \, 3\rangle}{[1 2]}\frac{[1 3]}{\langle 1 3\rangle} \pm \frac{[ n\!+\! 2\, 3]}{\langle 1 2\rangle }\frac{\langle 2\, n\!+\!2\rangle}{[2\, n\!+\!2]}\right)\, , 
  \>
where again the upper sign corresponds to the symmetric case, which will be the object most directly comparable to the simultaneous double-soft limit, and the lower sign to the antisymmetric case. At subleading order we find for the symmetric/antisymmetric case
 \<
   {\tt s/aCSL}^{(1)}(n+2, 1^{+}, 2^{-}, 3) &=&
   \pm \frac{1}{2} \frac{1}{[n\!+\!2 \, 2]^2}\frac{[n\!+\!2\, 1]}{\langle n\!+\!2 \, 1 \rangle} + \frac{1}{2} \frac{1}{\langle 1 3\rangle^2}\frac{\langle 2 3\rangle}{[2 3]}\nn\\
   & &+ \frac{1}{2} \frac{\langle n\!+\!2\,  3\rangle \langle 1 2\rangle \pm \langle n\!+\!2\, 2\rangle \langle 1 3\rangle}{[2 3]\langle n\!+\!2\, 1\rangle \langle 1 2\rangle \langle 1 3\rangle }\lambda_{2}^{\alpha}\frac{
   \partial}{\partial \lambda_3^{\alpha}}\nn\\
   & &+ \frac{1}{2} \frac{[ n\!+\!2\,  2] [1 3]  \pm [ n\!+\!2\,  3] [ 1 2]}{[2 3]\langle n\!+\!2\,  1\rangle [1 2] [2 3]} {\tilde \lambda}_{1}^{\dot \alpha}\frac{
   \partial}{\partial {\tilde \lambda}_{n+2}^{\dot \alpha}}\nn\\
   & &+ \frac{1}{2} \frac{[1 3]}{[1 2][2 3]}\frac{{\tilde \lambda}_{1}^{\dot \alpha}}{\langle 1 3\rangle}
   \frac{\partial}{\partial {\tilde \lambda}_3^{\dot \alpha}}\pm \frac{1}{2} \frac{\langle n\!+\!2\,  2\rangle}{\langle n\!+\!2\,  1\rangle\langle 1 2 \rangle}\frac{\lambda^\alpha_{2}}{[ n\!+\! 2\, 2]} \frac{
   \partial}{\partial \lambda_{n+2}^{\alpha}}~.
 \>
  As before we find some simplifications for the
 antisymmetric combination of consecutive limits, 
  \<
   {\tt aCSL}^{(1)}(n+2, 1^{+}, 2^{-}, 3)&=& \frac{1}{2}\frac{1}{\langle 1 3\rangle^2}\frac{\langle 2 3\rangle}{[2 3]} - \frac{1}{2} \frac{1}{[n\!+\!2 \, 2]^2}\frac{[n\!+\!2\,  1]}{\langle n\!+\!2 \, 1\rangle}\nn\\
     & &+\frac{1}{2}\frac{{\tilde \lambda}_{1}^{\dot \alpha}}{[1 2]}\biggbrk{\frac{1}{ [ n\!+\!2\,  2] } \frac{[ n\!+\!2\,  1] }{\langle n\!+\!2\,  1\rangle}\frac{
   \partial}{\partial {\tilde \lambda}_{n+2}^{\dot \alpha}} + \frac{1}{[2 3]} \frac{[1 3]}{\langle 1 3\rangle } \frac{\partial}{\partial {\tilde \lambda}_3^{\dot \alpha}}}\nn\\
   & &-\frac{1}{2}\frac{\lambda^\alpha_{2}}{\langle 1 2 \rangle }\biggbrk{ \frac{1}{\langle n\!+\! 2\, 1\rangle} \frac{\langle n\!+\!2\,  2\rangle}{[n\!+\!2\,  2]} \frac{
   \partial}{\partial \lambda_{n+2}^{\alpha}}+\frac{1}{\langle 1 3\rangle}\frac{\langle 2 3\rangle}{[2 3]} \frac{
   \partial}{\partial \lambda_{3}^{\alpha}}}~.
 \>

\paragraph{Gravity.} We can repeat the above considerations for the gravitational case and similarly define the consecutive soft limit factor ${\tt CSL}(1^{h_1}, 2^{h_2})$ as
first taking particle $2$ to be soft and then $1$. If both gravitons have positive helicity we find at leading order
\<
{\tt CSL}^{(0)}(1^{+}, 2^{+})&=&S^{(0)}(2^+)S^{(0)}(1^+) = \frac{1}{\langle 1 2\rangle^4  }\sum_{a,b\neq1, 2}^{n+2}\frac{[2 a][1 b]}{\langle 2 a\rangle \langle 1 b\rangle}\langle 1 a\rangle^2 \langle 2 b\rangle^2\ , 
\>
where we have used the freedom to choose the reference spinors in the two soft factors separately. Specifically, we chose the two reference spinors in $S^{(0)}(2^+)$ to be $\lambda_{1}$ and those in $S^{(0)}(1^+)$ to be $\lambda_{2}$. This makes the symmetry in particles 1 and 2 manifest, such that
\beq
\label{acslgpp}
{\tt aCSL}^{(0)}(1^{+}, 2^{+}) \ = \ 0 
\ . 
\eeq 
We see that the consecutive soft limit naturally involves a double sum over the external legs. 

At the next order we have 
\<
{\tt CSL}^{(1)}(1^{+}, 2^{+})&=&S^{(0)}(2^+)S^{(1)}(1^+)+S^{(1)}(2^+)S^{(0)}(1^+)~.
\>
Once again there will in principle be contact terms, which involve only a single sum over external legs, specifically 
\<
S^{(1)}(2^+)S^{(0)}(1^+)=\frac{1}{2}\sum_{a\neq 1,2}\frac{[2 a][1 2]}{\langle 2 a\rangle \langle 1 2\rangle}\frac{\langle x' a\rangle\langle y' a\rangle  }{\langle x' 1\rangle\langle y' 1\rangle}
+\mbox{non-contact terms}\, ,
\>
where $x'$ and $y'$ denote the reference spinors for the first particle. Choosing as above $\lambda_{x'}=\lambda_{y'}=\lambda_{2}$, we see that this contact term vanishes by momentum conservation. The complete subleading consecutive soft term is thus
\<
{\tt CSL}^{(1)}(1^{+}, 2^{+})&=&\frac{1}{\langle 1 2\rangle^3}\sum_{a, b\neq 1,2} \frac{[2 a][1 b]}{\langle 2 a\rangle\langle 1 b\rangle}\langle 1 a\rangle\langle 2 b\rangle\bigg[
\langle 2 b\rangle{\tilde \lambda}_{2}^{\dot \alpha} \frac{\partial}{\partial{\tilde \lambda}_{a}^{\dot \alpha}}- \langle 1 a\rangle {\tilde \lambda}_{1}^{\dot \alpha} \frac{\partial}{\partial{\tilde \lambda}_{b}^{\dot \alpha}}\bigg]~.
\>
Due to the absence of the contact term the expression is naturally symmetric in $q_1$ and $q_2$ and so ${\tt aCSL}^{(1)}(1^{+}, 2^{+})$ also vanishes. 

For the case where the first particle has positive helicity but the second has negative we find, for the same choice of reference spinors and to leading order, 
\<
\label{consgpgm}
{\tt CSL}^{(0)}(1^{+}, 2^{-})&=&\frac{1}{\langle 1 2\rangle^2 [1 2]^2 }\sum_{a,b\neq1, 2}^{n+2}\frac{\langle 2 a\rangle [1 b]}{[ 2 a] \langle 1 b\rangle}[ 1 a]^2 \langle 2 b\rangle^2~.
\>
A benefit of this choice of reference spinors is that it makes manifest that the order of soft limits does not matter, i.e. 
\beq
\label{acslgpm}
{\tt aCSL}^{(0)}(1^{+}, 2^{-})=0
\ . 
\eeq 
At subleading order we have, after taking the symmetric combination of soft limits,
\<
{\tt sCSL}^{(1)}(1^{+}, 2^{-})&=&
\frac{1}{2 \langle12\rangle[12]} 
 \sum_{a\neq 1, 2} \frac{[1 a]^2 \langle 2 a\rangle^2}{\langle 1 a\rangle^2  [2 a]^2} \langle a | q_{1{2}} | a]\nn\\
& & +
\frac{1}{\langle 1 2\rangle^2 [1 2]}\sum_{a, b\neq 1,2} \frac{\langle 2 a\rangle [1 b]}{[ 2 a]\langle 1 b\rangle}\bigg[
\langle 2 b\rangle^2  [ 1 a] { \lambda}_{2}^{ \alpha} \frac{\partial}{\partial{ \lambda}_{a}^{\alpha}}- \langle 1 a\rangle^2 [2 b] { \lambda}_{1}^{ \alpha} \frac{\partial}{\partial{ \lambda}_{b}^{ \alpha}}\bigg]~.
\>
We can of course continue to the sub-subleading terms, ${\tt CSL}^{(2)}$, however as the explicit expressions are involved we relegate
them to Appendix \ref{app:ss_csl}. However it is worth nothing that the sub-subleading terms involve a double contact term which 
has the same scaling as ${\tt CSL}^{(1)}$. If we consider the symmetrized version it has the form
\<
\left. {\tt sCSL}^{(2)}\right|_{dc}=\frac{1}{2[12]\vev{12}} \sum_{a\neq 1, 2} \biggbrk{ \frac{[1a]\vev{2a}^4}{\vev{1a}^3}+\frac{\vev{2a}[1a]^4}{[2a]^3} } ~,
\>
which should be combined with with $\left. {\tt sCSL}^{(1)}\right|_c$ to give
\<
\frac{1}{2 \langle12\rangle[12]} 
 \sum_{a\neq 1, 2} \frac{[1a]^3\vev{2a}^3}{\vev{1a}[2a]}\bigg[\frac{1}{\vev{a1}[1a]}\left(1-\frac{\vev{a2}[2a]}{\vev{a1}[1a]}\right)+\frac{1}{\vev{a2}[2a]}\left(1-\frac{\vev{a1}[1a]}{\vev{a2}[2a]}\right)\bigg]~.
\>
Notably for ${\tt CSL}^{(1)}$ the contact term does not vanish and so we have a non-trivial antisymmetric combination
\<
\label{aCSLc+-}
{\tt aCSL}^{(1)}(1^{+}, 2^{-})&=& \frac{1}{2 \langle12\rangle[12]} 
 \sum_{a\neq 1, 2} \frac{[1 a]^2 \langle 2 a\rangle^2}{\langle 1 a\rangle^2  [2 a]^2} \langle a | q_{1\bar{2}} | a] \ , 
\>
where $q_{1\bar 2}=q_1-q_2$. This term is more local than might be naively expected, rather in having the form of a single sum over
hard legs it is more like a single-soft factor.

\section{Simultaneous double-soft gluon limits}
\label{soft-gluons}

\subsection{Summary of results}

In this section we turn to the simultaneous double-soft limits, where we set $\delta_1 = \delta_2 =: \delta$ and expand the amplitude in powers of $\delta$. Correspondingly, we define the ``double-soft limit factor'' by
\be
{\tt DSL}(n\!+\!2,1^{h_{1}},2^{h_{2}},3)\, A_{n}(3,\ldots, n\!+\!2) = \lim_{\delta\to0} A_{n+2}(\delta q_1^{h_{1}}, \delta q_2^{h_{2}},3,\ldots , n\!+\!2) \, ,
\label{DSLgluonformula}
\ee
where the corresponding expansion of the double-soft function in $\delta$ is,
\<
{\tt DSL}(n+2,1^{h_1}, 2^{h_2},3) = \sum_{i} \delta^{i-2} {\tt DSL}^{(i)}(n+2, 1^{h_1}, 2^{h_2}, 3)~.
\>
The leading double-soft factor for the $1^+2^+$ helicity configuration may be straightforwardly derived from the formula of the generic MHV gluon amplitude. For the $1^+2^-$ helicity case, it is sufficient to consider the split-helicity six-point amplitude $A_{6}(5^{+},6^{+},1^{+},2^{-},3^{-},4^{-})$.%
\footnote{The explicit expression for the latter amplitude can be found e.g.~in Exercise 2.2 of \cite{Henn:2014yza}.} 
The results are
\begin{align}
\label{DSLpp}
{\tt DSL}^{(0)}(n\! + \! 2,{1}^{+},{2}^{+},3)&= \frac{\vev{n\! + \! 2\,  3}}{\vev{n\! + \! 2{1}}\vev{{1}{2}}\vev{{2}3}}=S^{(0)} (n+2, 1^+, 2)  \  S^{(0)} (n+2, 2^+, 3)  \, , 
\\ 
\label{DSLpm}
{\tt DSL}^{(0)}(n\! + \! 2,1^{+},2^{-},3)&= \frac{1}{\langle n\! + \! 2|q_{12}|3]}\,
\bigg[ \frac{1}{2 k_{n\! + \! 2}\cdot q_{12}}\, \frac{\bev{n\! + \! 2\, 3} \vev{n\! + \! 2 \, 2}^{3}}{\vev{12}\vev{n\! + \! 2\, 1}} - \frac{1}{2 k_{3}\cdot q_{12}}\, 
\frac{\vev{n\! + \! 2\,  3}\bev{31}^{3}}{\bev{12}\bev{23}}
\bigg]\, , 
\end{align}
where
\beq
\qquad q_{12}:=q_{1}+q_{2} \, .
\eeq
These formulae were tested numerically using {\tt S@M} \cite{Maitre:2007jq} and {\tt GGT} 
\cite{Dixon:2010ik,Schuster:2013aya} for a wide range of MHV, NMHV and NNMHV amplitudes from lengths 6 through 14.
Importantly these formulae do not have a ``local'' expression, i.e.~they may not be written
as a sum over a density depending on the two soft and one hard leg. Both hard legs are entangled.
In the next section we will present a derivation of \eqref{DSLpp} and \eqref{DSLpm} based on BCFW recursion relations \cite{Britto:2004ap,Britto:2005fq}. 

The sub-leading corrections to \eqref{DSLpp} and \eqref{DSLpm} are also computed via BCFW recursions in the following section and we present the results below:

\begin{align}
\hspace{-0.5cm}
{\tt DSL}^{(1)}(n+2,1^{+},2^{+},3)&=S^{(0)} (n+2, 1^+, 2)S^{(1)} (n+2, 2^+, 3)+S^{(0)} (1,2^+, 3)S^{(1)} (n+2, 1^+, 3),\nn \\\label{DSlsubpp}
\\
\hspace{-0.5cm}{\tt DSL}^{(1)}(n+2,1^{+},2^{-},3)&=S^{(0)} (n+2, 1^+, 2)S^{(1)} (n+2, 2^-, 3)+S^{(0)} (3, 2^-, 1)S^{(1)} (n+2, 1^+, 3)\nn\\
&+\frac{\vev{23}\bev{13}}{\bev{32}\vev{12}}\frac{1}{2p_3\cdot q_{12}}\laDla(2,3)+\frac{\vev{n+2\, 2}\bev{2\,n+2}}{\bev{n+2\, 1}\vev{12}}\frac{1}{2p_{n+2}\cdot q_{12}}\laDla(2,n+2)\nn\\
&+\frac{\bev{n+2\,1}\vev{2\,n+2}}{\vev{1\, n+2}\bev{21}}\frac{1}{2p_{n+2}\cdot q_{12}}\tlaDtla(1,n+2)+\frac{\bev{31}\vev{32}}{\vev{13}\bev{21}}\frac{1}{2p_3\cdot q_{12}}\tlaDtla(1,3)\nn\\
&+{\tt DSL}^{(1)}(n+2,1^{+},2^{-},3)\vert_{c},
\label{DSLsubpm}
\end{align}
 where,
\begin{align}
{\tt DSL}^{(1)}(n+2,1^{+},2^{-},3)\vert_{c}&= \frac{\vev{n+2\,2}^2 \bev{1\, n+2}}{\vev{n+2\, 1}}\frac{1}{(2p_{n+2}\cdot q_{12})^2}+\frac{\bev{31}^2 \vev{23}}{\bev{32}}\frac{1}{(2p_3\cdot q_{12})^2}.
\label{DSLsubpmcon}
\end{align}

It is interesting to note that the results for both the leading and the sub-leading simultaneous double-soft function for the $1^+2^+$ gluons are same as the consecutive soft limits in the previous section. However,  the case with the $1^+2^-$ is considerably different than the consecutive soft limits scenario and we get new terms especially the last two lines in \eqref{DSLsubpm} look like some deformation of $S^{(1)} (n+2, 2^-, 3)$ and $S^{(1)} (n+2, 1^+, 3)$ respectively, due to the double-soft limit.  Moreover, we also have the contact terms\eqref{DSLsubpmcon} which are absent for the previous case.
\subsection{Derivation from BCFW recursion relations}

In the application of the BCFW recursion relation we consider a $\langle 1 2]$ shift, i.e.\ a holomorphic shift of momentum of the first soft particle and an anti-holomorphic shift of the momentum of the second one, specifically we define
\beq
\label{bcfw-s}
\hat\lambda_{1} := \lambda_{1} + z \lambda_{2} \, , \qquad \hat{\tilde \lambda}_{2} := \tilde\lambda_{2} - z \tilde\lambda_{1} \ . 
\eeq
The first observation to make is that generic BCFW diagrams with the soft legs belonging to the left or right $A_{n>3}$ amplitudes are subleading in the soft limit.%
\footnote{This observation was made in \cite{ArkaniHamed:2008gz} in relation to the study of a double-soft scalar limit. There, the  relevant diagrams turned out to be those involving a four-point functions, and are indeed finite. }
This is because the shifted momentum of a soft leg turns hard through the shift in a generic BCFW decomposition. The exception is when any of the two soft legs belongs to a three-point amplitude. Thus nicely, there are two special diagrams to consider, namely those where either one of the two soft particles belongs to a three-point amplitude. 
In the following we consider separately two cases: $1^+ 2^+$ and $1^+ 2^-$. 


\subsubsection*{The $\mathbf{1^+2^+}$ {\bf case}.}
\begin{figure}[t]
\centering
\includegraphics[width=0.5\linewidth]{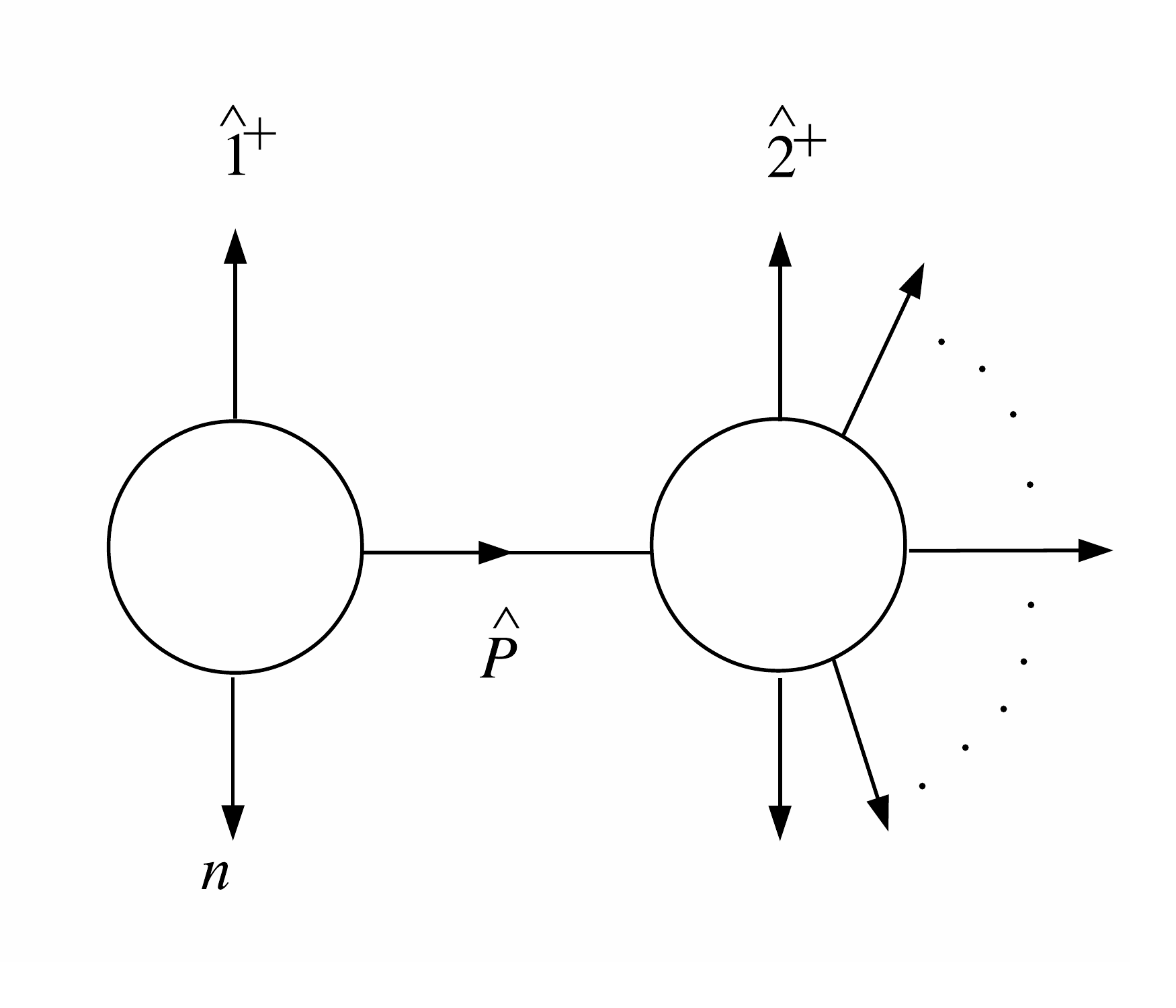}
\caption{\it The first BCFW diagram contributing to the double-soft factor. The amplitude on the left-hand side is $\overline{\rm MHV}$.}
\label{fig1}
\end{figure}

There are two special   BCFW diagrams to consider. The first one is shown  in Figure \ref{fig1}, where the three-point amplitude sits on the left with the external legs $\hat{1}$ and $n\!+\!2$ (with the remaining legs $2, \ldots, 
n\!+\!1$ on the right-hand side). A second diagram has the three-point amplitude  on the right-hand side, with external   legs $\hat{2}$ and $3$.  
In the first diagram, the three-point amplitude has the $\overline{\rm MHV}$ helicity configuration because of our choice of $\langle 1 2]$ shifts.  One easily finds that the solution to $\langle \hat{1} 2 \rangle =0$ is 
\beq 
\label{zast}
z_\ast\  = \ - {\langle 1 \, n\!+\!2 \rangle \over   \langle 2 \,  n\!+\!2 \rangle}
\ , 
\eeq
and note that $z_\ast$ stays constant as particles $1$ and $2$ become soft. 
One also finds 
\beq
\hat{\lambda}_1 = -  {\langle 12 \rangle \over \langle 2 \,  n\!+\!2 \rangle }\, \lambda_{n+2}
 \ , 
 \eeq
as well as 
\beq
\lambda_{\hat{P}} \tilde\lambda_{\hat{P}}   \ = \la_{n+2}(\tla_{n+2}+\frac{\vev{12}}{\vev{n+2\,2}}\tla_1)
\label{Pshiftdiag1}
\eeq
If we were taking just particle $2$ soft, the shifted momentum $\hat{2}$ would remain hard. However we are taking a simultaneous double-soft limit where  
both particles $1$ and $2$ are becoming soft, and as a consequence the momentum $\hat{2}$ becomes soft as well, see \eqref{bcfw-s} and \eqref{zast}. 
Thus, we can take a soft limit also on the amplitude on the right-hand side. The diagram in consideration then becomes
\beqa
\label{softpp}
&&
A_3 \big( (n\!+\!2)^+, \hat{1}^+, \hat{P}^-\big) {1\over (q_1 + p_{n+2})^2} \,
A_{n} (\hat 2^{+}, \ldots ,\hat{P})
\ , 
\eeqa
Using the explicit expression for the three-point anti-MHV amplitude and the shifts derived earlier, and  also \eqref{Pshiftdiag1},
we may rewrite the right-hand subamplitude in the above with the soft shifted leg $\hat 2$ as
\beqa
\hspace{-0.61cm}
A_{n} (\hat 2^{+}, \ldots ,p_{n+2} + \delta
\tfrac{\vev{12}}{\vev{n+2\, 2}}\, |n+2\rangle\, [1|)
&=
e^{\delta \frac{\vev{12}}{\vev{n+2\, 2}}\, [1 \partial_{n+2}]}\, 
\Bigl(\frac{1}{\delta}\, S^{(0)}(n+2,\hat 2^{+},3) + S^{(1)}(n+2,\hat 2^{+},3)\nn\\
& +\delta\, S^{(2)}(n+2,\hat 2^{+},3)\Bigr)\,  A_{n}(3,\ldots)\, ,
\eeqa
where, we define,
\beq
[i \partial_{j}]:=\tlaDtla(i,j)
\eeq
From this expressions all relevant leading and subleading contributions to the simultaneous
double-soft factor 
\begin{align}
{\tt DSL}(n+2,1^+, 2^+,3)&= \frac{A_3 \big( (n\!+\!2)^+, \hat{1}^+, \hat{P}^-\big) }{(q_1 + p_{n+2})^2} \,
\nonumber \\
&\, e^{\delta \frac{\vev{12}}{\vev{n+2\,2}}\, [1 \partial_{n+2}]}\, 
\Bigl ( \frac{1}{\delta}\, S^{(0)}(n+2,\hat 2^{+},3) + S^{(1)}(n+2,\hat 2^{+},3) +\delta\, S^{(2)}(n+2,\hat 2^{+},3)\Bigr )\,
\label{master++gluon}
\end{align}
may be extracted. Expanding the above expression in $\delta$, at leading order we get,
\be
{\tt DSL}^{(0)}(n+2,1^+,2^+,3)={  \langle n\!+\!2\,  3 \rangle  \over \langle n\!+\!2\, 1 \rangle \langle 12 \rangle \langle 23 \rangle} \, .
\label{gluon++LO}
\ee
For the sake of definiteness we have 
considered particle $n\!+\!2$ to have positive helicity; a  similar analysis can be performed for the case where $n\!+\!2$ has negative helicity, and leads to the very same conclusions.
Note that this contribution \eqref{softpp}   diverges as $1/\delta^2 $ if we scale the  soft momenta as $q_i \to \delta q_i$, with $i=1,2$. 
 \begin{figure}[t]
\centering
\includegraphics[width=0.5\linewidth]{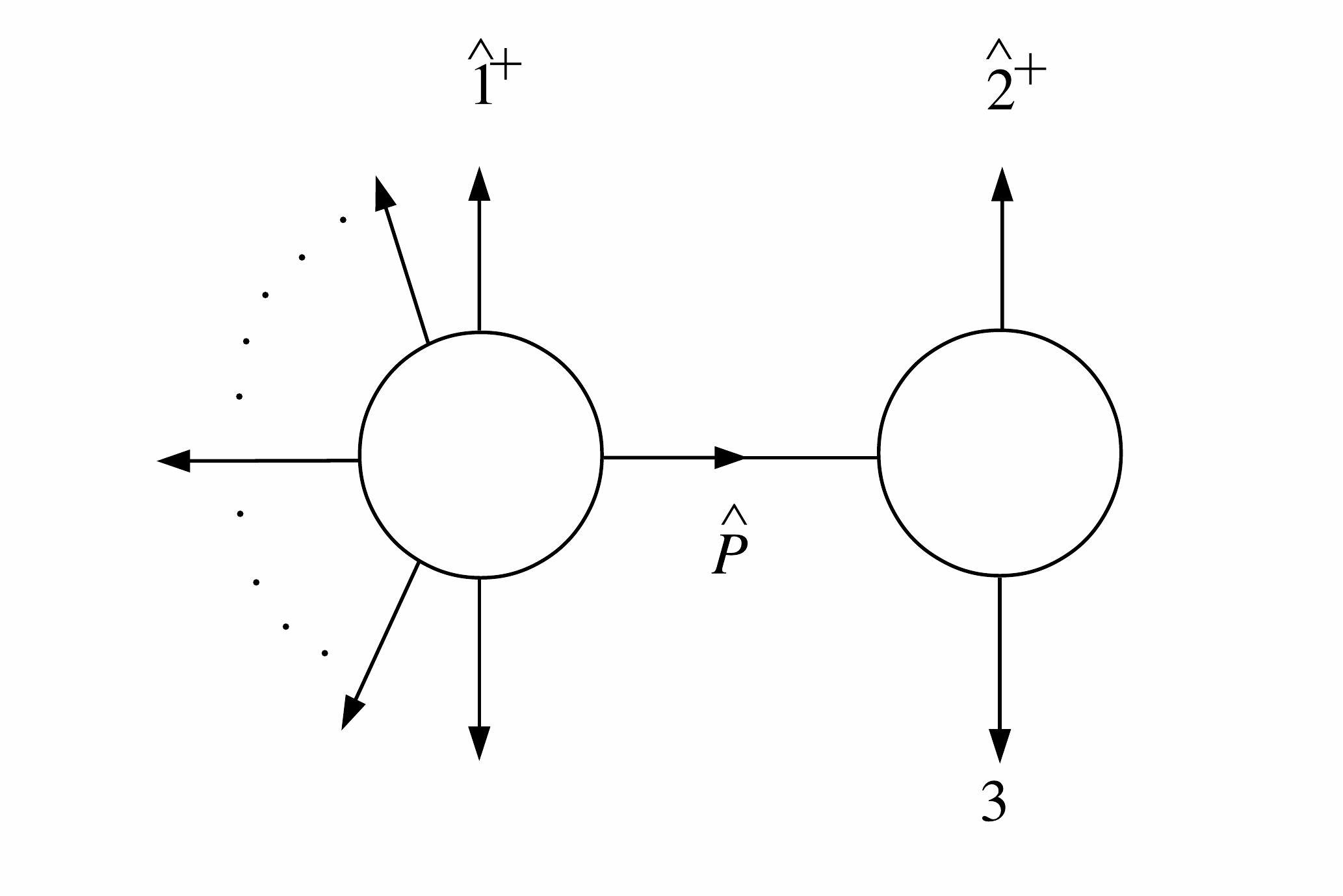}
\caption{\it The second BCFW diagram contributing to the double-soft factor. The three-point amplitude is {\rm MHV}. For the case where gluon $2$ has positive helicity we find that this diagram is subleading compared to that in Figure \ref{fig1} and can be discarded; while when  $2$ has negative helicity this diagram is as leading as Figure \ref{fig1}.}
\label{fig2}
\end{figure}
 There still is another diagram to compute, shown in Figure \ref{fig2} but we now show that it is in fact subleading. In this diagram, the amplitude on the right-hand side is a three-point amplitude with particles $\hat{2}^+$,   $3$ and $\hat{P}$. If particle $3$ has positive helicity, then the three-point amplitude is $\overline{\rm MHV}$ and hence vanishes because of our shifts. Thus we have to consider only the case when particle $3$ has negative helicity. In this case we have 
 the diagram is 
 \beq
 A_3 ( \hat{2}^+,    3^-, \hat{P}^-)
 {1\over (q_2 + p_3)^2}  \, A_{n+1}( \hat{1}^+, \hat{P}^+, 4, \ldots, (n\!+\!2)^+)\ . 
 \eeq
 Similarly to the case discussed earlier, the crucial point is that leg $\hat{1}^+$ is becoming soft as the momenta $1$ and $2$ go soft. The diagram then becomes 
 \beq
 \label{zzz}
 A_3 ( \hat{2}^+ ,    3^ -, \hat{P}^-)
 {1\over (q_2 + p_3)^2} \, S^{(0)}( n\!+\!2, \hat{1}^+, \hat{P}) \, A_{n}(\hat{P}^+, 4, \ldots, (n\!+\!2)^+)\ , 
 \eeq
 and note that $A_{n}\big(\hat{P}^+, 4, \ldots, (n\!+\!2)^+\big) \to A_{n}\big(3+, 4, \ldots, (n\!+\!2)^+\big)$ in the soft limit. We can now evaluate the prefactor in \eqref{zzz} using that, for this diagram, $z_\ast = [23] / [13]$ and 
 \beq 
 \tilde\lambda_2 = \tilde\lambda_3 {[12] \over [13]} \, , 
 \qquad 
 \lambda_{\hat{P}} \tilde{\lambda}_{\hat{P}} \   =  (\lambda_{3} + \frac{\bev{12}}{\bev{13}}\, \lambda_{2}) \tilde\lambda_3   \  .
 \eeq
In the soft limit we find 
\beq
A_3 ( \hat{2}^+ ,   3^ -, \hat{P}^-)
 {1\over (q_2 + p_3)^2} \, S^{(0)}( n\!+\!2, \hat{1}^+, \hat{P})  \to { [12]^3\over [23][31]}\,  {1\over p_3 \cdot q_{12}} \, { \langle n\!+\!2 \, 3 \rangle \over \langle n\!+\!2 |\,  q_{12} \,   | 3]} 
  \ , 
  \eeq
  which is finite under the scaling $q_i \to \delta q_i$, with $i=1,2$, and hence subleading with respect to \eqref{softpp}. In conclusion, we find for the double-soft factor for soft gluons  $1^+ 2^+$:
\beq
A_{n+2} (1^+, 2^+, 3, \ldots , n) \ \to  \ {\tt DSL}(n\!+\!2,1^{+},2^{+},3) \, A_n ( 3, \ldots, n+2)
\ , 
\eeq
with 
\beq
{\tt DSL}^{(0)}(n\!+\!2,1^{+},2^{+},3)\, = \, \frac{\vev{n\!+\!2\, 3}}{\vev{n\!+\!2\, 1}\vev{12}\vev{23}} \ , 
\eeq
which agrees with \eqref{DSLpp}.

A comment is in order here. We observe that the BCFW diagram in Figure \ref{fig1} is precisely the diagram contributing to the single-soft gluon limit identified originally in \cite{Cachazo:2014fwa} and later studied in \cite{Casali:2014xpa} for Yang-Mills. In the simultaneous double-soft limit, particle $\hat{2}$ also becomes soft thanks to the shifts, and hence we can approximate the BCFW diagram by further extracting a single-soft function for a gluon with soft, shifted momentum $\hat{2}$: 
\beq
A_{n+2} (1^+, 2^+, 3, \ldots , n+2) \ \to \     S^{(0)} (n+2, 1^+, 2) \, S^{(0)} (n+2, \hat{2}^+, 3) \, A_{n} (3, \ldots, n\!+\!2)   \ . 
\eeq
Moreover, because of our $\langle 1 2]$ shifts and the holomorphicity of the soft factor for a single positive-helicity gluon, we have that $S^{(0)} (n+2, \hat{2}^+, 3) = S^{(0)} (n+2, 2^+, 3)$, thus 
\beq
 {\tt DSL}^{(0)}(n\!+\!2,1^{+},2^{+},3) \ = \ S^{(0)} (n+2, 1^+, 2) \, S^{(0)} (n+2, 2^+, 3)
 \ . 
 \eeq
In fact, we can immediately see that a consecutive limit, where particles $1$ and $2$ are taken soft one after the other  (as opposed to our simultaneous double-soft limit) would give the same result. Indeed one would get
\beqa
A_{n+2} (1^+, 2^+, 3, \ldots , n+2) &\to & S^{(0)} (n+2, 1^+, 2)  \,  A_{n+1} (2, \ldots, n\!+\!2)  
\nonumber \\ \cr
&\to &   
 S^{(0)} (n+2, 1^+, 2)  \  S^{(0)} (n+2, 2^+, 3)  \, A_{n} (3, \ldots, n\!+\!2)
\ , 
\eeqa
in other words at the leading order, the simultaneous double-soft factor for same-helicity soft gluons is nothing but the consecutive soft limit given by the product of two single soft gluon factors. 

Now, we present the subleading term in the expansion of \eqref{master++gluon}, which scales as $\delta^{-1}$,
\begin{align}
{\tt DSL}^{(1)}(n+2,1^{+},2^{+},3)&= -\frac{\vev{n+2\, 2}}{\vev{n+2\, 1}\vev{12}}\bigg(\frac{1}{\vev{23}}\tlaDtla(2,3)+\frac{1}{\vev{n+2\, 2}}\tlaDtla(2,n+2)\bigg)\nn\\
&-\frac{\vev{1 3}}{\vev{12}\vev{23}}\bigg(\frac{1}{\vev{13}}\tlaDtla(1,3)+\frac{1}{\vev{n+2\, 1}}\tlaDtla(1,n+2)\bigg)
\label{dsl++gluonNLO}
\end{align}
and the previous equation can be further simplified in terms of leading and subleading terms of single-soft functions as,
\begin{align}
\hspace{-0.5cm}
{\tt DSL}^{(1)}(n+2,1^{+},2^{+},3)&=S^{(0)} (n+2, 1^+, 2)S^{(1)} (n+2, 2^+, 3)+S^{(0)} (1,2^+, 3)S^{(1)} (n+2, 1^+, 3).
\label{dsl++gluonNLOsimple}
\end{align}
Note that this contribution was only from the first type of BCFW diagram discussed above, the second type was finite already at the leading order so it again does not contribute to the subleading term here.

\subsubsection*{The $\mathbf{1^+2^-}$ {\bf case}.}
We turn again  to the  two diagrams considered in the previous case. However, we will see that this time they are  both  leading. Consider  the first diagram. The only difference compared to \eqref{softpp} is the soft factor, which now has to be replaced with $S^{(0)} ( \hat{P}, \hat{2}^-, 3)$ since particle $2$ has now negative helicity.  We use the same shifts, and make use of the results 
\beq
\hat{\tilde\lambda}_2 \ =  \ { q_{12} \,  \, | n\!+\!2 \rangle \over \langle 2 \,  n\!+\!2 \rangle}  \, , \qquad
\tilde{\lambda}_{\hat{P}}   \ =  \ 
{ (q_1 + p_{n+2}) |2 \rangle \over \langle 2 \, n\!+\!2 \rangle }  \ . 
\eeq
Using this, we evaluate the soft factor as 
\beq
{ [ \hat{P} 3] \over [ \hat{P} \hat{2} ] [ \hat{2} 3] } \    \to    \     { [3 |\,  n\!+\!2 \,  | 2 \rangle  \over  [3 | \, q_{12}\,  | \, n\!+\!2 \, \rangle  }\ {\langle n\!+\!2 \, 2 \rangle \over 2 p_{n+2} \cdot \, q_{12} } 
\ . 
\eeq
The diagram in consideration is then quickly seen to give 
\beq
\label{df1}
{ [3\, n\!+\!2]\, \langle n\!+\!2\, 2 \rangle^3 \over \langle 12 \rangle \langle n\!+\!2\,  1 \rangle } { 1\over  [3 | \, q_{12} \, | \, n\!+\!2 \, \rangle} {1\over 2 p_{n+2} \cdot\, q_{12} } \, A_{n} (3, \ldots , n\!+\!2)
\ . 
\eeq
Next we move to the second diagram. Again, in principle one has to  distinguish  two cases depending on the helicity of particle 3, but it is easy seen that such cases  turn out to give the same result. For the sake of definiteness we illustrate  the situation where particle $3$ has positive helicity. We obtain 
\beq
{ \langle \hat{P} 2 \rangle^3 \over \langle 23\rangle \langle 3 \hat{P}\rangle}    {1\over \langle 23\rangle [32] } \, S^{(0)} (n\!+\!2, \hat{1}^+, \hat{P})    \, A_{n} (\hat{P}, 4, \ldots, n\!+\!2)
\ . 
\eeq
Using 
\beq
\tilde\lambda_{\hat{P}} = {[1 | (q_2 + p_3) \over [13]} \, , \qquad  \hat{\lambda}_1 = {q_{12}\,  |3 ] \over [13]}\ , 
\eeq
we easily see that this contribution gives, to leading order in the soft momenta,  
\beq
\label{df2}
{ \langle n\!+\!2\, 3 \rangle [13]^3  \over [12] [23]} {1\over  \langle n\!+\!2 | \, q_{12}\,  | 3 ] }\,   {1\over 2 p_3 \cdot q_{12} } \  A_{n} (3, 4, \ldots, n\!+\!2)
 \ . 
 \eeq
Putting together \eqref{df1} and \eqref{df2} one obtains for the double-soft factor for soft gluons  $1^+ 2^-$:
\beq
A_{n+2} (1^+, 2^-, 3, \ldots , n) \ \to  \ {\tt DSL}(n\!+\!2,1^{+},2^{-},3) \, A_n ( 3, \ldots, n+2)
\ , 
\eeq
with 
\beq
{\tt DSL}^{(0)}(n\!+\!2,1^{+},2^{-},3) \,= \, {1\over  \langle n\!+\!2 | \, q_{12}\,  | 3 ] }\bigg[
 {1\over 2 p_{n+2} \cdot\, q_{12} }\, 
{ [n\!+\!2\, 3]\, \langle n\!+\!2\, 2 \rangle^3 \over \langle 12 \rangle \langle n\!+\!2\,  1 \rangle }    \ - \ 
{1\over 2 p_3 \cdot q_{12} } \, { \langle n\!+\!2\, 3 \rangle [31]^3  \over [12] [23]} \bigg]
  \, , 
  \label{eq:p1m2dslcorrect}
 \eeq
 which agrees with  \eqref{DSLpm}. 

As already observed earlier, we comment that the diagrams in Figure \ref{fig1} and \ref{fig2} are precisely the BCFW diagrams which would contribute to the single-soft gluon limit when either gluon $1$ or $2$ are taken soft, respectively. Thus, the result we find for the double-soft limit has the structure 
\beq
{\tt DSL}^{(0)}(n\!+\!2,1^{+},2^{-},3)    \ = \  S^{(0)} (1^+)\, S^{(0)} (\hat{2}^- ) \ + \  S^{(0)} (2^-)\, S^{(0)} (\hat{1}^+ )
\ , 
\eeq
with the two contributions arising from Figure \ref{fig1} and \ref{fig2}, respectively. The situation however is less trivial than in the case where the two soft gluons had the same helicity, and the double-soft factor is not the product of two single-soft factors. 

Now, following the steps for the case of $\{1^+,2^+\}$ gluons, we can derive the subleading corrections to the double-soft function. However, unlike the previous case here we will have to take into account the contribution from both the BCFW diagrams \ref{fig1} and \ref{fig2} . 
\begin{align}
{\tt DSL}^{(1)}(n+2,1^{+},2^{-},3)&=\frac{\bev{3\,n+2}\vev{n+2\,2}^3}{\vev{n+2\, 1}\vev{12}\langle n+2|q_{12}|3](2p_{n+2}\cdot q_{12})}\bigg(\frac{-(2p_{n+2}\cdot q_{12})}{\bev{3\,n+2}\vev{n+2\,2}}\laDla(2,3)\nn\\
&+\frac{\langle n+2|q_{12}|3]}{\bev{3n+2}\vev{n+2\,2}}\laDla(2,n+2)-\frac{\vev{12}}{\vev{n+2\, 2}}\tlaDtla(1,n)\bigg)\nn\\
&+\frac{\vev{n+2\, 3}\bev{13}^3}{\bev{32}\bev{21}\langle n+2|q_{12}|3](2p_{3}\cdot q_{12})}\bigg(\frac{-(2p_3\cdot q_{12})}{\bev{13}\vev{n+2\,3}}\tlaDtla(1,n+2)\nn\\
&+\frac{\langle n+2|q_{12}|3]}{\bev{13}\vev{n+2\,3}}\tlaDtla(1,3)-\frac{\bev{21}}{\bev{13}}\laDla(2,3)\bigg)+{\tt DSL}^{(1)}(n+2,1^{+},2^{-},3)\vert_{c},
\label{DSL+-gluonNLOa}
\end{align}
where contribution to the subleading terms coming from the contact terms, i.e. the ones with no derivative operator,  and these are given by
\begin{align}
{\tt DSL}^{(1)}(n+2,1^{+},2^{-},3)\vert_{c}&=\frac{\vev{n+2\,2}^2 \bev{1\, n+2}}{\vev{n+2\, 1}}\frac{1}{(2p_{n+2}\cdot q_{12})^2}+\frac{\bev{31}^2 \vev{23}}{\bev{32}}\frac{1}{(2p_3\cdot q_{12})^2}.
\label{DSL+-gluoncontact}
\end{align}
 We note that the above equation can be simplified further as,
 \begin{align}
\hspace{-0.5cm}{\tt DSL}^{(1)}(n+2,1^{+},2^{-},3)&=S^{(0)} (n+2, 1^+, 2)S^{(1)} (n+2, 2^-, 3)+S^{(0)} (3, 2^-, 1)S^{(1)} (n+2, 1^+, 3)\nn\\
&+\frac{\vev{23}\bev{13}}{\bev{32}\vev{12}}\frac{1}{(2p_3\cdot q_{12})}\laDla(2,3)+\frac{\vev{n+2\, 2}\bev{2\, n+2}}{\bev{n+2 \,1}\vev{12}}\frac{1}{(2p_{n+2}\cdot q_{12})}\laDla(2,n+2)\nn\\
&+\frac{\bev{n+2\,1}\vev{2n+2}}{\vev{1 \,n+2}\bev{21}}\frac{1}{(2p_{n+2}\cdot q_{12})}\tlaDtla(1,n+2)+\frac{\bev{31}\vev{32}}{\vev{13}\bev{21}}\frac{1}{(2p_3\cdot q_{12})}\tlaDtla(1,3)\nn\\
&+{\tt DSL}^{(1)}(n+2,1^{+},2^{-},3)\vert_{c}.
\label{DSL+-gluonNLOsimple}
\end{align}

\section{Simultaneous double-soft graviton limits}

\subsection{Summary of results}

The analysis of the double-soft limit of gravitons  in terms of the BCFW recursion relations for General Relativity 
\cite{Bedford:2005yy,Cachazo:2005ca}
is entirely similar to that of  gluons described in the previous section. As before, we scale the momenta of the soft particles as  $q_i \to \delta q_i$, $i=1,2$. 
The main result here is that,  at leading order in $\delta$ and for both choices of  helicities of the gravitons becoming soft,   the double-soft factor is nothing but the product of two single-soft particles (and we recall that the order in which the gravitons are taken soft is immaterial to this order, see \eqref{acslgpp} and \eqref{acslgpm}). 
Specifically, we define the graviton double-soft limit factor by
\be
{\tt DSL}(1^{h_{1}},2^{h_{2}})\, M_{n}(3, \ldots, n\!+\!2)
= \lim_{\delta \to 0} M_{n+2}(\delta q_1^{h_{1}}, \delta q_2^{h_{2}}, 3, \ldots, n\!+\!2)
\ee 
and find
\begin{align}
\label{GDSLpp}
 {\tt DSL}^{(0)}({1}^{h_{1}}, {2}^{h_{2}}) \ & =   \ S^{(0)} ({1}^{h_{1}}) S^{(0)} 
 ({2}^{h_{2}})    \\
 {\tt DSL}^{(1)}({1}^{h_{1}}, {2}^{h_{2}}) \ & =   \ S^{(0)} ({1}^{h_{1}}) S^{(1)} ({2}^{h_{2}})+   S^{(0)} ({2}^{h_{2}}) S^{(1)} ({1}^{h_{1}})
 +{\tt DSL}^{(1)}({1}^{h_{1}}, {2}^{h_{2}})|_{c}
 \ ,
\end{align}
where $S^{(i)}(s^\pm)$ are the single-soft factors for graviton $s^\pm$ given in \eqref{singlesoftgrp}. The contact term at subleading order, ${\tt DSL}^{(1)}({1}^{h_{1}}, {2}^{h_{2}})|_{c}$, vanishes for identical helicities $h_{1}=h_{2}$ of the soft gravitons and takes the form
\begin{equation}
{\tt DSL}^{(1)}({1}^{+}, {2}^{-})|_{c}=
\frac{1}{ q_{12}^2}\sum_{a\neq 1,2} 
\frac{\bev{1a}^{3}\vev{2a}^3}{\vev{1a}\bev{2a}}\frac{1}{2\, p_a\cdot q_{12}}
\, ,
\end{equation}
in the mixed helicity case.
 Note that both double-soft factors diverge at leading order as $1/ \delta^2$. 
Differences to the consecutive soft-limit appear only in the contact term at subleading order
$1/ \delta$ in the mixed helicity case. 

\begin{figure}[t]
\centering
\includegraphics[width=0.5\linewidth]{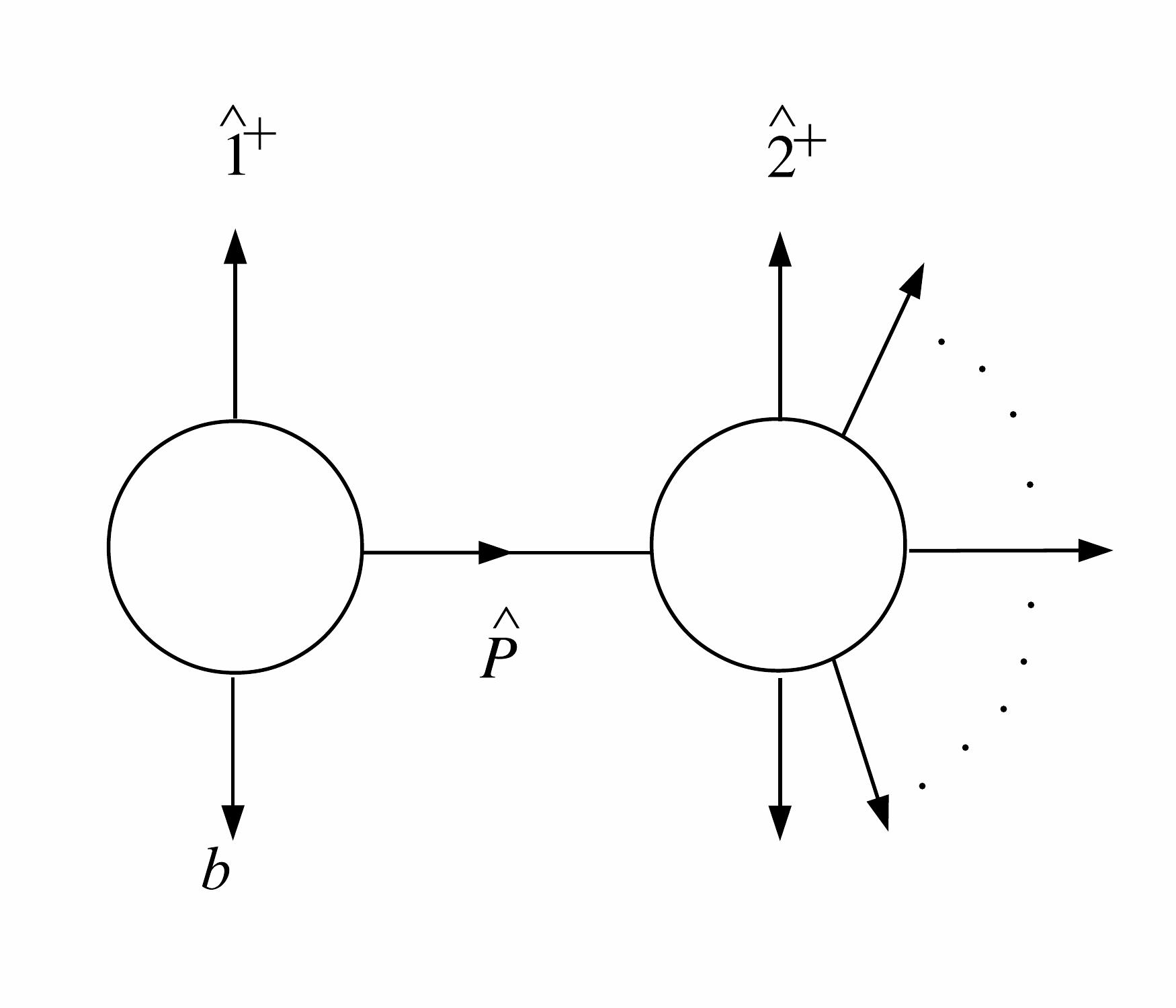}
\caption{\it The first class of BCFW diagrams contributing to the double-soft factor for two gravitons. The amplitude on the left-hand side is $\overline{\rm MHV}$, and one has to sum over all possible choices of the graviton $b$.}
\label{fig3}
\end{figure}

\subsection{Derivation from the BCFW recursion relation}

As for the case of gluons, we distinguish two cases depending on whether the two gravitons becoming soft have the same or opposite helicities. We outline below  the main steps of the derivations.

\subsubsection*{The $1^+2^+$ case}

The first relevant class of diagram is shown in Figure \ref{fig3}, where $b$ can be any of the $n$ hard particles. For the sake of definiteness we illustrate the case where $b$ has positive helicity; the case where $b$ has negative helicity leads to an identical result. Using the fact that the momentum $\hat{q}_2$ is becoming soft we can write this diagram as 
\beq
\label{softpp-grav}
\hspace{-0.4cm}
M_3 ( b^+, \hat{1}^+, \hat{P}^-) {1\over (q_1 + p_b)^2} \, M_{n} (\hat 2^{+}, \hat{P}, \ldots )  \ , 
\eeq
where $S^{(0)} (s^+)$ is given in \eqref{singlesoftgrp},  
and $x$ and $y$ denote two arbitrary reference spinors. Using the explicit expression for the three-point anti-MHV amplitude and the shifts derived earlier, and that $\hat{P}=p_{b} + \delta
\frac{\vev{1b}}{\vev{2b}}\, |b\rangle\, [1|$ 
we may rewrite the last term in the above with the soft shifted leg $\hat 2$ as
\beq
M_{n} (\hat 2^{+}, p_{b} + \delta
\tfrac{\vev{1b}}{\vev{2b}}\, |b\rangle\, [1|, \ldots )
=
e^{\delta \frac{\vev{1b}}{\vev{2b}}\, [1 \partial_{b}]}\, 
\biggl( \frac{1}{\delta}\, S^{(0)}(\hat 2^{+}) + S^{(1)}(\hat 2^{+}) +\delta\, S^{(2)}(\hat 2^{+}) \biggr) \,  M_{n}(b,\ldots)\, .
\eeq
From this expressions all relevant leading and subleading contributions to the simultaneous
soft factor may be extracted:
\be
{\tt DSL}(1^+, 2^+)= \frac{M_3 ( b^+, \hat{1}^+, \hat{P}^-) }{(q_1 + p_b)^2} \,
e^{\delta \frac{\vev{1b}}{\vev{2b}}\, [1 \partial_{b}]}\, 
\biggl( \frac{1}{\delta}\, S^{(0)}(\hat 2^{+}) + S^{(1)}(\hat 2^{+}) +\delta\, S^{(2)}(\hat 2^{+})\biggr) \, .
\label{master++}
\ee
At leading order we find
\beq
{\tt DSL}^{(0)}(1^+, 2^+)\ M_{n} (b, \ldots ) \ , 
\eeq
with 
\beqa
\label{DSLppgrav}
 {\tt DSL}^{(0)}(1^+, 2^+) & = & {1\over \langle 12 \rangle^2 }\sum_{b\neq 1,2} 
 { [b1] \langle b2 \rangle^2 \over \langle 1 b \rangle} S^{(0)} ( \hat{2} ) 
 \nonumber \\
 &=& 
 {1\over \langle 1 2 \rangle^2 }\sum_{a, b\neq 1,2} {[b1] \langle b2 \rangle \over \langle 1b \rangle } 
 {\langle b | \, q_{12}\,  | a] \over \langle 2a \rangle  } { \langle x a \rangle \langle y a \rangle  \over \langle x 2 \rangle \langle y2 \rangle}\ .
\eeqa
The expression \eqref{DSLppgrav} is   symmetric in the two soft particles, $1$ and $2$, although not manifestly. Furthermore, it turns out using total momentum conservation that 
\beq
{\tt DSL}^{(0)}(1^+, 2^+) = S^{(0)} (1^+) \, S^{(0)} (2^+)
\ , 
\eeq
i.e.~the double-soft factor for gravitons with the same helicity is the product of two single-soft factors.  
Again it is not a local expression, in the sense explained in  Section \ref{soft-gluons}. 

One can also work out the first subleading contribution to the double-soft limit. 
The result reads for the non-contact term
\begin{align}
{\tt DSL}^{(1)}(1^+, 2^+)|_{nc} = \frac{1}{\vev{12}^2} \sum_{a,b\neq 1,2} 
\frac{\bev{b1}\vev{b2}}{\vev{1b}}\, \frac{\langle b|q_{12}|a]}{\vev{2a}}\, \biggl[ &
\frac 1 2 \biggl( \frac{\vev{xa}}{\vev{x2}}+ \frac{\vev{ya}}{\vev{y2}} \biggr ) \,
\biggl( \tilde \la_{2}^{\da}\, \frac{\partial}{\partial \tla_{a}^{\da}} 
+\frac{\vev{1b}}{\vev{2b}}\, \tilde \la_{1}^{\da}\, \frac{\partial}{\partial \tla_{a}^{\da}}\biggr)
\nn\\ 
& +\frac{\vev{xa}\vev{ya}\vev{12}}{\vev{x2}\vev{y2}\vev{b2}}\, \tilde \la_{1}^{\da}\, \frac{\partial}{\partial \tla_{b}^{\da}}\, \biggr]
\end{align}
Making the gauge choice $\la_{x}=\la_{y}=\la_{1}$ to make contact to the discussion in 
section \ref{consec-sec} we find
\begin{align} 
{\tt DSL}^{(1)}(1^+, 2^+)|_{nc} &= \frac{1}{\vev{12}^{3}}
\sum_{a,b\neq 1,2} 
\frac{\bev{b1}\vev{b2}}{\vev{1b}}\, \frac{\langle b|q_{12}|a]\,\vev{1a}}{\vev{2a}}\, \biggl[ 
\tilde \la_{2}^{\da}\, \frac{\partial}{\partial \tla_{a}^{\da}} 
+ \frac{\vev{1b}}{\vev{2b}}\, \tilde \la_{1}^{\da}\, \frac{\partial}{\partial \tla_{a}^{\da}}
 - \frac{\vev{1a}}{\vev{2b}}
 \, \tilde \la_{1}^{\da}\, \frac{\partial}{\partial \tla_{b}^{\da}} \,  
\biggr] \, . 
\end{align}
In fact the middle term vanishes by momentum conservation $\sum_{b} |b]\langle b|=0$. The structure may be further reduced by splitting up the $\langle b|q_{1} + q_{2}|a]$ factor and
using momentum conservation and the Lorentz invariance $\sum_{b}[1b]\, [1\tilde\partial_{b}]\cA =0$. This lets us rewrite this double-soft factor as
\[
{\tt DSL}^{(1)}(1^+, 2^+)|_{nc} =  S^{(0)}(1^{+})\,S^{(1)}(2^{+})
+S^{(0)}(2^{+})\, S^{(1)}(1^{+}) \, .
\]
We also get a contact term contribution to the above subleading factor when the derivative
operator $[1\partial_{b}]$ in the exponential in (\ref{master++}) hits the leading
soft function $S^{(0)}(\hat 2^{+})$, 
\begin{align}
{\tt DSL}^{(1)}(1^+, 2^+)\vert_{c} &= \frac{\bev{12}}{\vev{12}^{3}}\,
\langle 1 | \sum_{b\neq 1,2} p_{b}|1] =0 
\, .
\end{align}

\begin{figure}[t]
\centering
\includegraphics[width=0.5\linewidth]{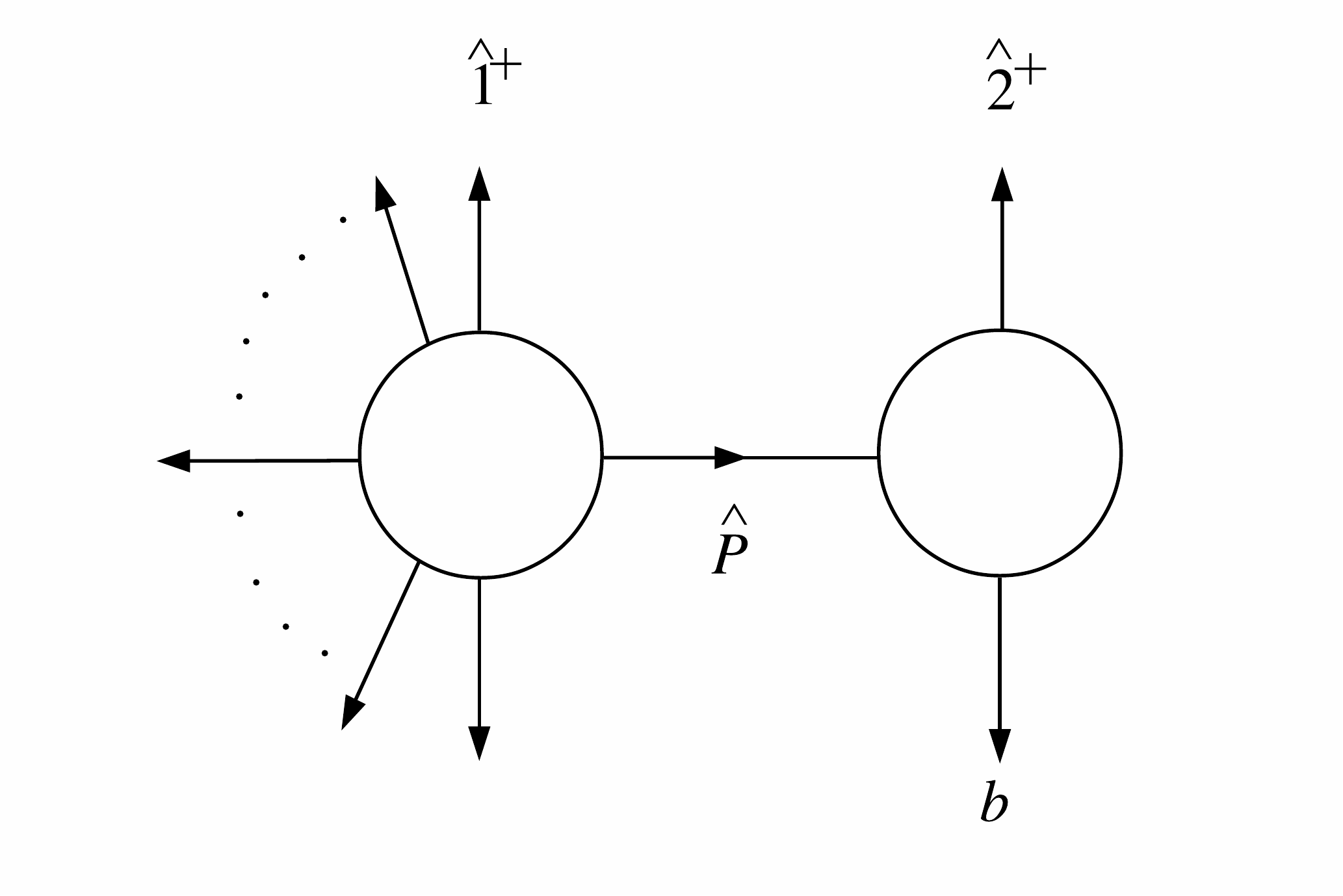}
\caption{\it The second class of BCFW diagram contributing to the double-soft graviton factor. The three-point amplitude is {\rm MHV} and one has to sum over all possible choices of graviton $b$. Similarly to the gluon case, this diagram contributes only when  graviton $2$ has negative helicity.}
\label{fig4}
\end{figure}

As for the case of soft gluons, we have to consider another diagram which is however vanishing as we take the two particles soft. This diagram is depicted in Figure \ref{fig4}. 
A short calculation shows that the contribution of this diagram is at the leading order in 
$\delta$
\beq
\left( {\langle \hat{P} 3 \rangle^3 \over \langle \hat{P} 2 \rangle   \langle 2 3 \rangle }\right)^2 {1\over \langle 2 b \rangle [ b2 ] } \, S^{(0)} ( \hat{1}^+) \ = \ { [12]^6 \over [13]^2 [23]^2 } \, S^{(0)} ( \hat{1}^+)
\ , 
\eeq
times an $n$-point amplitude. This quantity is immediately seen to vanish as we take the momenta of particles $1$ and $2$ soft and thus irrelevant at the first three leading orders.
Similarly, one also convinces oneself that the generic BCFW diagram with $n>3$ point amplitudes to the right or left is finite in the soft limit and therefore not contributing to the considered leading orders. As soon as diagrams of this type start contributing the universality is lost and there is no double-soft factor.

\subsubsection*{The $1^+2^-$ case}

The analysis of this case proceeds in a very similar way as for gluons. Again there are two diagrams contributing, depicted in Figures \ref{fig3} and \ref{fig4}. The calculations of these diagrams is straightforward and involves the soft factors $S (\hat{2}^-)$ and $S (\hat{1}^+)$, respectively. These soft factors are given by,%
\footnote{Recall that we are using a $\langle 1 2]$ shift, which explains the various hatted quantities in \eqref{11} and \eqref{22}.} 
\beqa
\label{11}
S^{(0)} (\hat{2}^-)  & = & \sum_{a\neq 1,2}  { \langle 2 a \rangle [xa] [ya] \over [ \hat{2} a] [x\hat{2} ] [y\hat{2} ] } 
\, , \qquad  S^{(1)} (\hat{2}^-)  = \frac 12  \sum_{a\neq 1,2} 
\frac{\vev{2a}}{\bev{\hat 2 a}}\, \left ( \frac{\bev{xa}}{\bev{x\hat 2}} + 
\frac{\bev{ya}}{\bev{y\hat 2}} \right ) \, \vev{2 \partial_{a}}
\\
\label{22}
S^{(0)} (\hat{1}^+)  & = & \sum_{a\neq 1,2}  { [1 a]  \langle xa\rangle  \langle ya\rangle \over \langle \hat{1} a\rangle  \langle x \hat{1} \rangle \langle  y\hat{1} \rangle  } \, , 
 \qquad  S^{(1)} (\hat{1}^+)  = \frac 12  \sum_{a\neq 1,2} 
\frac{\bev{1a}}{\vev{\hat 1 a}}\, \left ( \frac{\vev{xa}}{\vev{x\hat 1}} + 
\frac{\vev{ya}}{\vev{y\hat 1}} \right ) \, \bev{1 \partial_{a}}
\eeqa
where 
\beq
\hat{\tilde{\lambda}}_2 = { q_{12}\,  | b \rangle \over \langle 2 b \rangle }  \ , 
\eeq
for the first recursive diagram, and 
\beq
\hat{\lambda}_1 \ = \ { q_{12}\, | b ] \over [ 1 b ] }
\ , 
\eeq
for the second one. It is particularly convenient to choose $\tilde\lambda_x = \tilde\lambda_y = \tilde\lambda_1$ and $\lambda_x = \lambda_y = \lambda_2$, for the first and second diagram, respectively. Doing so, we obtain from  the first diagram 
\beq
\label{a1}
\frac{1}{\delta}\, \frac{\vev{2b}^{2}\, \bev{b1}}{\vev{12}^{2}\, \vev{1b}}\,
e^{\delta \frac{\vev{12}}{\vev{b2}}\, \bev{1\partial_{b}}}\, \Bigl\{
\frac{1}{\delta}S^{(0)}(
\hat{2}^{-}) + S^{(1)}(\hat{2}^{-}) \, \Bigr \}\, M_{n}(3,\ldots, n+2)\, ,
 \eeq
while, for the second, 
\beq
\label{a2}
\frac{1}{\delta}\, \frac{\vev{2b}\, \bev{b1}^{2}}{\bev{12}^{2}\, \bev{b2}}\,
e^{\delta \frac{\bev{12}}{\bev{1b}}\, \vev{2\partial_{b}}}\, \Bigl\{
\frac{1}{\delta}S^{(0)}(\hat{1}^{+}) + S^{(1)}(\hat{1}^{+}) \, \Bigr \}\, M_{n}(3,\ldots, n+2)\,  .
 \eeq
The double-soft factor   for soft gravitons $1^+ 2^-$ is obtained by summing the two contributions in \eqref{a1} and \eqref{a2}. At leading order we find
\beq
\label{DSLpmgrav}
{\tt DSL}^{(0)}({1}^{+}, {2}^{-}) \ = \ 
{1\over q_{12}^4} \sum_{a, b\neq 1,2}\bigg[ 
 { \langle 2b \rangle^3 [1a]^2 [1b] \langle 2a \rangle   \over \langle 1b \rangle\,  \langle b | \, q_{12}\,  | a ] }  \ +  \ 
{ [1b]^3 \langle 2a \rangle^2 \langle 2b\rangle [1a]  \over [2b]\,  [b | \, q_{12}\,  | a \rangle} \bigg]
\ .
\eeq
In fact, we can easily combine the two terms  in \eqref{DSLpmgrav} and show that we just get the result of the consecutive limit discussed earlier in \eqref{consgpgm}.  To this end, in the second term in \eqref{DSLpmgrav} we relabel $a\leftrightarrow b$ and use 
\beq
{\langle 2 b \rangle \over \langle 1 b \rangle }+ {[1a] \over [2a]} = - {[ a | \, q_{12} \, | b \rangle \over \langle 1 b \rangle [ 2 a]}
\ . 
\eeq 
Hence we conclude that 
\beq
{\tt DSL}^{(0)}(1^+, 2^-) = S^{(0)} (1^+) \, S^{(0)} (2^-)
\ . 
\eeq
Working out the first subleading contribution to the double-soft limit
for the mixed helicity assignments from \eqref{a1} and \eqref{a2} one finds for the non-contact
terms
\begin{align}
{\tt DSL}^{(1)}(1^+, 2^-)|_{nc} &= \frac{1}{q_{12}^4}\sum_{a,b\neq 1,2} 
\, \frac{\bev{1a}^{2}\,  \bev{1b} \, \vev{2a}\, \vev{2b}^{2}}{\vev{b1}\, \bev{2a}}
\, 
\biggl ( \frac{\bev{12}}{\bev{1a}}\, \la^{\a}_{2}\, \frac{\partial}{\partial\la^{\a}_{a}}
-\frac{\vev{12}}{\vev{2b}}\, \tla^{\da}_{1}\, \frac{\partial}{\partial\tla^{\da}_{b}}\, \biggr )
\nn\\
& = \, S^{(0)} (1^+) \, S^{(1)} (2^-) +  S^{(0)} (2^-) \, S^{(1)} (1^+)
\, .
\end{align}
where the same gauge choices for the reference spinors as above were made. This subleading term also has a contribution from contact terms given by
\begin{align}
{\tt DSL}^{(1)}(1^+, 2^-)\vert_{c} &= \frac{1}{q_{12}^2}\sum_{b\neq 1,2} 
\,\biggl( \frac{\bev{1b}^{4}\,  \vev{2b}^3 }{\bev{b2}\, (2 p_b\cdot q_{12})^2}
\, +\frac{\bev{1b}^{3}\,  \vev{2b}^4 }{\vev{b1}\, (2 p_b\cdot q_{12})^2}\,\biggr )
\nn\\
& = \frac{1}{q_{12}^2}\sum_{b\neq 1,2} 
\,\frac{\bev{1b}^{3}\,  \vev{2b}^3 }{\bev{2b}\,\vev{1b}\, }\frac{1}{2 p_b\cdot q_{12}}
\, .
\end{align}
We hence see, that a difference to the consecutive double-soft limit appears at the subleading
order in the contact term above, cf. (\ref{aCSLc+-}).

\section{Double-soft scalars in $\mathcal{N}=4$ super Yang-Mills}

The emission of a single soft scalar in $\mathcal{N}=4 $ super Yang-Mills does not lead to any divergence -- the amplitude after a soft scalar has been emitted is in general finite. Thus, the consecutive limit where two scalars are taken soft is also finite and  not universal. It is then interesting that the simultaneous double-soft scalar limit does lead to a universal divergent structure, which can also be analysed using recursion relations. 

To begin it is useful to look at simple examples. We take two scalars in a singlet configuration, and consider the amplitudes $A (1_{\phi_{12}}, 2_{\phi_{34}}, g_3, g_4, g_5)$, where the helicities of the gluons $( g_3, g_4, g_5)$ are a permutation of $(- - +)$. It is then easy to extract the double-soft limit: 
\beq
\label{ex}
A (1_{\phi_{12}}, 2_{\phi_{34}}, g_3, g_4, g_5)  \  \to \   \frac{[23][1 5] \langle 5 3\rangle}{ s_{125} s_{123} [12]} \ A(g_3, g_4, g_5)
\ . 
\eeq
Note that the prefactor appearing in this equation is divergent in the double-soft limit. In the following we wish to derive such kind of  behaviour from a recursion relation. 
One direct approach is to perform the supersymmetric generalisation of the $\langle 12]$-shift used in previous sections:
\<
\label{bcfw-ss}
\hat\lambda_{1} &:=& \lambda_{1} + z \lambda_{2} \, , \qquad \hat{\tilde \lambda}_{2} := \tilde\lambda_{2} - z \tilde\lambda_{1} \ ,\qquad {\hat \eta}_2=\eta_2-z \eta_1~.
\>
As in the bosonic case there are two special BCFW diagrams to consider: Figure \ref{fig1}, where the three-point amplitude sits on the left with the external legs $\hat 1$ and 
$n+2$ and Figure \ref{fig2} with the three-point amplitude on the right-hand side with external legs $\hat 2$ and $3$ (where now particles 1 and 2 are scalars). If we take the holomorphic limit discussed in 
Appendix \ref{app:susy_soft} for both particle $1$ and $2$ we will find the supersymmetric generalisation of the 
bosonic $1^+ 2^+$ case. Instead we will consider taking the holomorphic limit of particle $1$ and the antiholomorphic limit of particle $2$ which is the supersymmetric generalisation of the
 $1^+2^-$ case; as in that case we find contributions from both BCFW diagrams. 
 The calculation is essentialy identical to the bosonic case and so we will omit the details. The contribution from Figure \ref{fig1} is
 \<
\int d^4\eta_P~ A_3^{\overline{\rm MHV} }(n\!+\!2, \hat{1}, \hat{P})\frac{1}{\vev{1\, n\!+\! 2}[n\!+\! 2\, 1]} {\bar S}(-\hat{P}, \hat 2, 3)A_n(-\hat P, 3, \dots)\, , 
\>
where $A_3^{\overline{\rm MHV}}$ is the supersymmetric $\overline{\rm MHV}$ three-point amplitude and ${\bar S}(a, s, b)$ is the antiholomorphic
 soft factor described in Appendix \ref{app:susy_soft}. Performing the integrations over the internal Gra\ss mann parameters we can extract the
 contribution to the appropriate double-soft factor by examining the coefficient of the relevant $\eta$'s. For particle $1$ and $2$ being scalars in the
  singlet state, i.e. the coefficient of the $\eta_1^2\eta_2^2$ term, the leading order contribution is
  \<
{\tt DSL}_{\rm a}(n+2, 1_\phi, 2_\phi, 3)=\frac{\vev{n\!+\!2\, 2}[n\!+\!2\, 3]\vev{n\!+\!2\, 1}}{2 p_{n+2}\cdot q_{12} \vev{12} \langle n\!+\!2| q_{12}|3]}~.
\>
The contribution from Figure \ref{fig2} is
\<
\int d\eta_P~ S(n+2, \hat 1, \hat P) A_n(n+2, \hat{P}, \dots)\frac{1}{p^2_{23}}A_3^{\rm MHV}(\hat 2, 3, -\hat P)\, , 
\>
where now $S(a,s,b)$ is the holomorphic factor in Appendix \ref{app:susy_soft}. This diagram contributes to the singlet scalar double-soft coefficient the term
\<
{\tt DSL}_{\rm b}(n+2, 1_\phi, 2_\phi, 3)=-\frac{\vev{ n\!+\!2 \, 3} [31][32]}{2p_3 \cdot q_{12} \langle n\!+\!2| q_{12}|3][12]}~.
\>
To find the complete double soft factor we combine the two terms i.e.
\<
{\tt DSL}(n+2, 1_\phi, 2_\phi, 3)={\tt DSL}_{\rm a}(n+2, 1_\phi, 2_\phi, 3)+{\tt DSL}_{\rm b}(n+2, 1_\phi, 2_\phi, 3)~.
\>
 For the sake of illustration, we derive the result \eqref{ex} for the particular case of $(g_3, g_4, g_5) = (3^-, 4^-, 5^+)$, with the scalars in a flavour singlet configuration. 
 Due to the three-particle kinematics  we have 
 \<
 {\tilde \lambda}_3\propto  {\tilde \lambda}_4 \propto  {\tilde \lambda}_5 \, , 
 \>
 and hence 
 for this particular choice the contribution from ${\tt DSL}_{\rm a}$ is zero. Moreover we can exchange $|5]$ and $|3]$
 in the expression ${\tt DSL}_{\rm b}$ as the constants of proportionality cancel between the numerator and denominator,  hence 
 \beq
{\tt DSL}_{\rm b}(5, 1_\phi, 2_\phi, 3)=- {\langle  5 3\rangle  [3 1] [3 2]  
 \over \langle 3 | q_{12} | 3]  [1\, 2] \langle 5 | q_{12} | 3] } = {\langle  5 3\rangle   [51]  [2 3 ]
 \over \langle 3 | q_{12} | 3]  [1\, 2] \langle 5 | q_{12} | 5] }
 \ , 
 \eeq
in agreement with \eqref{ex} at leading order in the double-soft expansion.  

\begin{figure}[t]
\centering
\includegraphics[width=0.5\linewidth]{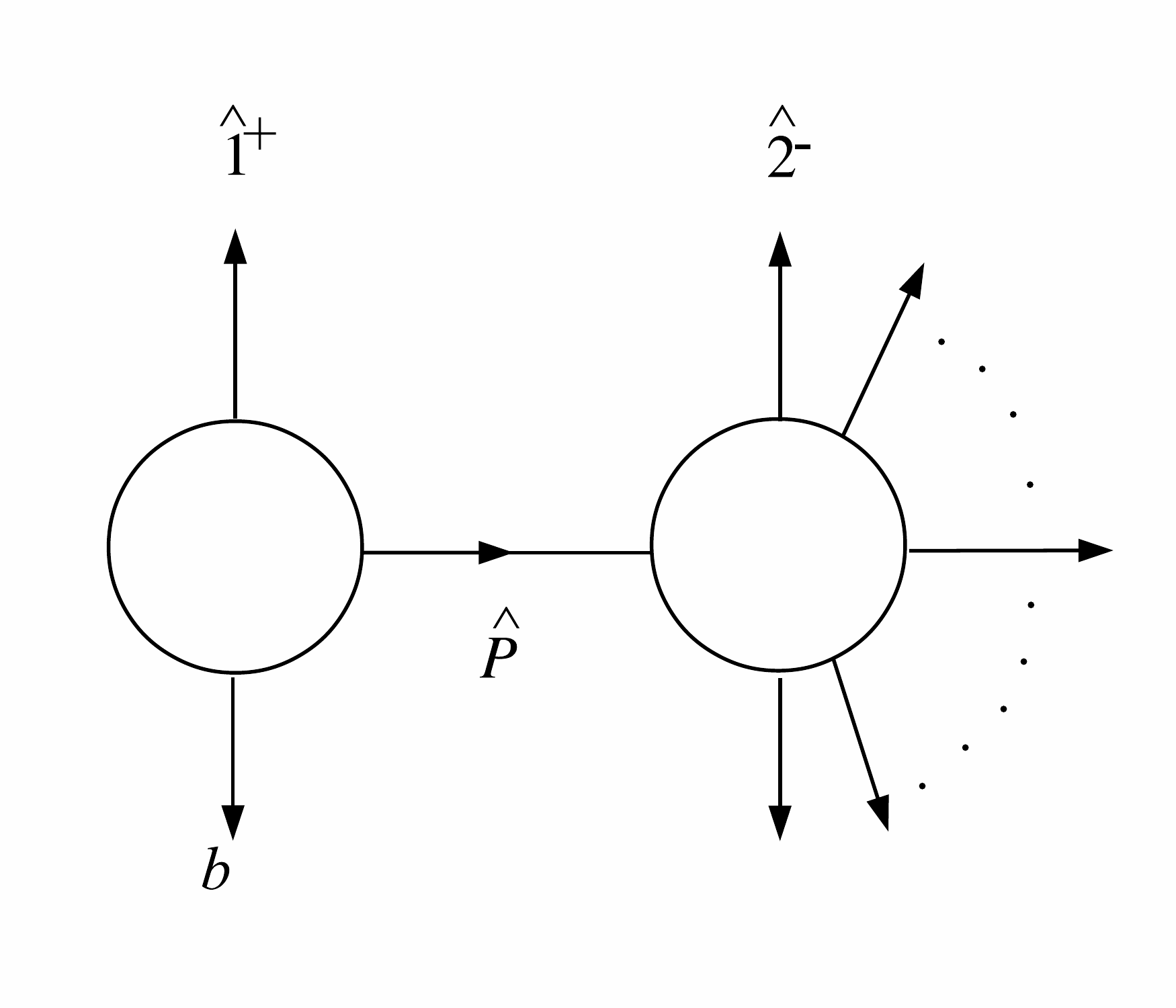}
\caption{\it The first BCFW diagram contributing to the double-soft scalar limit. }
\label{fig5}
\end{figure}

\begin{figure}[t]
\centering
\includegraphics[width=0.5\linewidth]{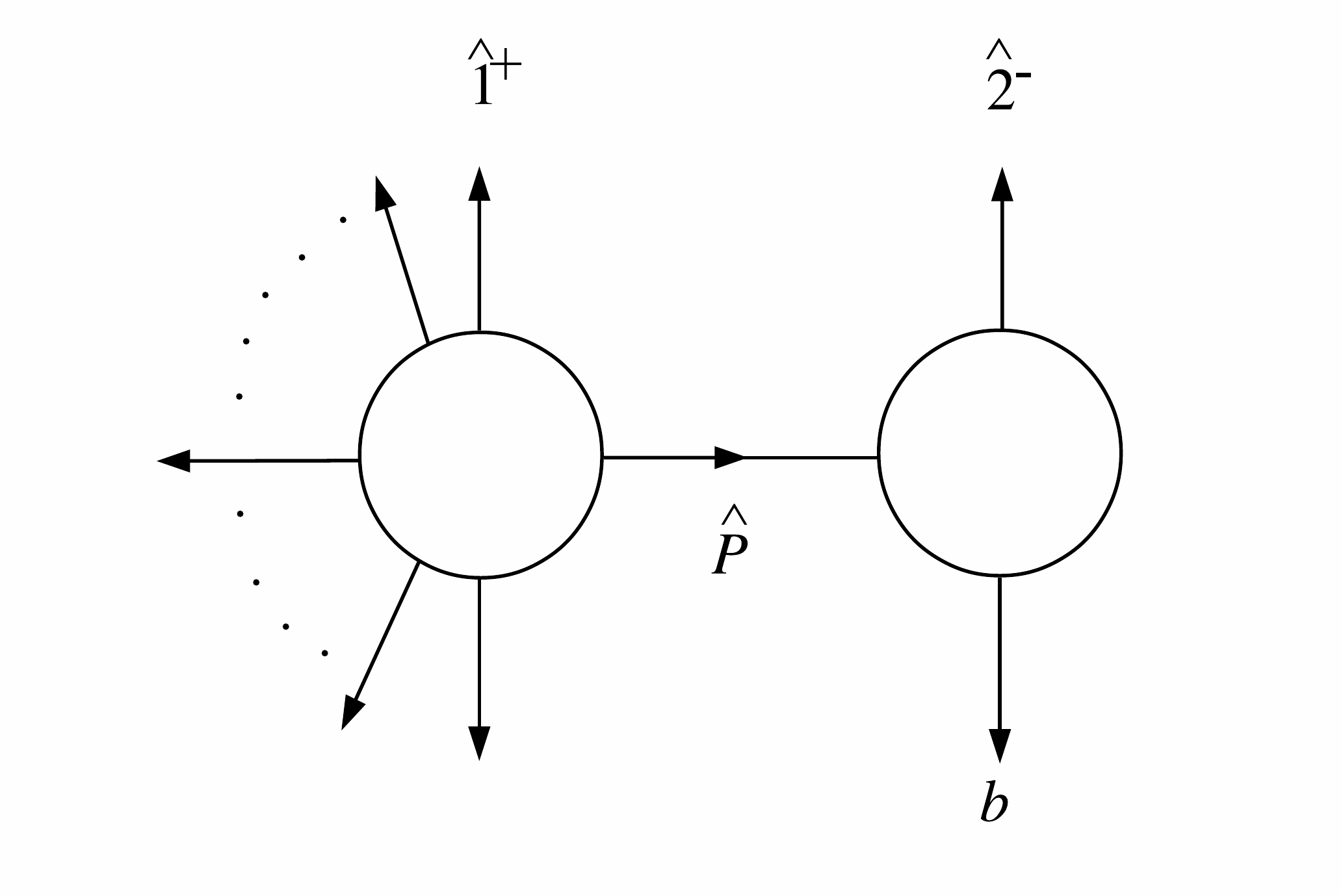}
\caption{\it The second BCFW diagram contributing to the double-soft scalar limit. This diagram does not contribute when the two scalars are in a flavour non-singlet configuration.}
\label{fig6}
\end{figure}

We can also re-derive this result from a different recursion relation, where we shift one of the two soft particles and one hard particle. Taking again the scalars in positions 1 and 2, we shift one of the scalars, say $2$, and an adjacent hard particle $3$, 
\beq
\lambda_{\hat{2 }} = \lambda_2 + z \lambda_3 \, , \quad 
\tilde{\lambda}_{\hat{3 }} =\tilde{\lambda}_3 - z \tilde{\lambda}_2 \, , \quad 
\eta_{\hat{3}} = \eta_3  - z \eta_2 \, . 
\eeq
There are two recursion diagrams to consider, shown in Figures \ref{fig5} and \ref{fig6}.  We begin discussing the first one, where we have  a four-point amplitude with both soft legs attached to it. To leading order in the soft parameter $\delta$,  the position of the pole in $z$ is
\beq
z_{\ast}  \ =\  \frac{2\, p_n \cdot q_{12}}{\langle 3\, n\!+\!2 \rangle [ 2\,  n\!+\!2 ] }  \ . 
\eeq 
The  BCFW diagram in Figure \ref{fig5} is then
\beq
\label{rec-susy}
A_{n+2} \ =\   \int\!d^4 \eta_{ \hat{P} } \
A_{4}(n+2, 1, \hat{2},  \hat{P} ) \frac{1}{P^2} A_{n} ( -\hat{P}, \hat{3}, \ldots ) \, ,
\eeq
where  $P^2 = (q_{12} + p_{n+2})^2\simeq 2 q_{12} \cdot p_{n+2}$,  and the four-point superamplitude  is explicitly given by 
\beq
A_{4}(1, \hat{2}, \hat{P} , n\!+\!2) \ 
= \ 
\frac{  \delta^{(8)}( \lambda_1 \eta_1 + \lambda_{\hat{2} } \eta_2 + \lambda_{\hat{P}} \eta_{\hat{P}} + \lambda_{n+2} \eta_{n+2})}{
\langle 1 \hat{2} \rangle \langle \hat{2} \hat{P} \rangle \langle \hat{P} n\!+\!2 \rangle \langle n\!+\!2\, 1 \rangle
}
\, .
\eeq
We can re-write the fermionic delta function as
\beqa
\delta^{(8)}( \lambda_1 \eta_1 + \lambda_{\hat{2} } \eta_2 + \lambda_{\hat{P}} \eta_{\hat{P}} + \lambda_{n+2} \eta_{n+2})  &=& \langle \hat{2} \hat{P} \rangle ^4 \, \delta^{(4)} \Big( \eta_{\hat{P}} + \eta_1 {\langle 1 \hat{2} \rangle \over \langle \hat{P} \hat{2}\rangle} + \eta_{n\!+\!2} {\langle n\!+\! 2 \, \, \hat{2}\rangle \over \langle\hat{P} \hat{2}\rangle}\Big)
\nonumber \\ && \delta^{(4)} \Big( \eta_2 + \eta_1 {\langle 1 \hat{P} \rangle \over \langle \hat{2}\hat{P} \rangle} + \eta_{n\!+\!2} {\langle n\!+\!2 \hat{P}\rangle \over \langle\hat{2} \hat{P}\rangle}\Big)\ , 
\eeqa
thus getting
\beq
\label{if}
{ \langle \hat{2} \hat{P} \rangle^3
\over 
\langle 1 \hat{2} \rangle \langle \hat{P} \, n\!+\!2 \rangle \langle n\!+\!2\, 1 \rangle}\, 
  \delta^{(4)} \Big( \eta_2 + \eta_1 {\langle 1 \hat{P} \rangle \over \langle \hat{2}\hat{P} \rangle} + \eta_{n\!+\!2} {\langle n\!+\!2 \hat{P}\rangle \over \langle\hat{2} \hat{P}\rangle}\Big)%
\, 
A_{n} ( -\hat{P}, \hat{3}, \ldots, n\!+\!1 )
\ , 
\eeq
where now $A_n$ is evaluated at 
\beq
  \eta_{\hat{P}} = 
- \eta_1  { \langle 1\hat{2}  \rangle \over \langle  \hat{P}  \hat{2}\rangle}
 - \eta_{n\!+\!2} {   \langle n\!+\!2\, \hat{2} \rangle \over \langle \hat{P} \hat{2} \rangle }
 \ . 
 \eeq 
One can also easily work out%
\footnote{The $\sim$ sign means that an equality holds at leading order in the double-soft limit.}
\begin{align}
\nonumber
\langle 1 \hat{P}\rangle & \sim \langle 1 \, n\!+\!2\rangle\, , &
\langle \hat{2}  \hat{P}\rangle & \sim {\langle 1\,  n\!+\!2 \rangle [ n\!+\!2 \, 1]  \over [n\!+\!2 \, 2]}\, , 
\\
\langle \hat{P} \, n\!+\!2 \rangle &\sim { [1\, 2]\langle 1 \, n\!+\!2\rangle   \over [n\!+\!2\, 2]}\, 
\ , 
&
\langle 1 \hat{2}\rangle & = { \langle n\!+\!2 \, 1\rangle \langle 3 | q_{12} | n\!+\!2]\over \langle 3\, n\!+\!2\rangle [n\!+\!2\, 2]}
\ , 
\label{relat}
\end{align}
so that \eqref{if} becomes
\beq
\label{ccc1}
{ [n\!+\!2\, 1]^3 \langle 3 \, n\!+\!2\rangle \over [n\!+\!2 \, 2] [1 2] \langle 3 | q_{12} | n\!+\!2]}
  \delta^{(4)} \Big( \eta_2 + \eta_1 {\langle 1 \hat{P} \rangle \over \langle \hat{2}\hat{P} \rangle} + \eta_{n\!+\!2} {\langle n\!+\!2 \, \hat{P}\rangle \over \langle\hat{2} \hat{P}\rangle}\Big)
\, 
A_{n} ( -\hat{P}, \hat{3}, \ldots, n\!+\!1 )\ .
\eeq

The second diagram is easily seen to contribute 
\beq
\label{ccc2}
{\langle 1 3 \rangle \over \langle 1 2 \rangle \langle 2 3 \rangle}  \, 
A_{n+1}    ( \{ - \lambda_1, \tilde\lambda_1 + \tilde\lambda_2 \tfrac{\langle 2 3 \rangle}{\langle 1 3 \rangle }, \eta_1 + \eta_2 \tfrac{\langle 2 3 \rangle}{\langle 1 3 \rangle}\}, \{ \lambda_3, \tilde\lambda_3 + \lambda_2 \tfrac{\langle 1 2 \rangle}{\langle 1 3 \rangle }, \eta_3 + \tfrac{\langle 1 2 \rangle}{\langle 1 3 \rangle }\eta_2\} \,,  \{4\} \ldots , \{ n+2\} )
\ , 
\eeq
where we notice that the prefactor is divergent only if we simultaneously make the momenta $q_1$ and $q_2$ soft. 

At this point we have to take components of (the sum of) \eqref{ccc1} and \eqref{ccc2}. One can distinguish two basic cases, namely whether the two scalars are in a singlet or non-singlet helicity configuration. In the latter case, only the recursion diagram in Figure \ref{fig5}, given by \eqref{ccc1},  contributes. 
 For the sake of illustration, we derive the result \eqref{ex} for the particular case of $(g_3, g_4, g_5) = (3^-, 4^-, 5^+)$, with the scalars in a flavour singlet configuration. For this particular choice, the diagram in Figure \ref{fig6} vanishes since the amplitude on the left-hand side would have to be {\rm MHV}, and thus vanishing given our choice of shifts. One is then left with the contribution from Figure \ref{fig5}, which is equal to 
 \beq
 {[5 1] [5 2] \langle 3 5 \rangle 
 \over \langle 3 4\rangle [34][1\, 2] \langle 3 | q_{12} | 5] } \, A_3 ( 3^-, 4^-, 5^+)
 \ , 
 \eeq
in agreement with \eqref{ex} at leading order in the double-soft expansion.

Next we discuss another  particularly simple situation,  where particle $3$ is a negative-helicity gluon, and we take the two scalars in a non-singlet flavour configuration. In this case the diagram of Figure \ref{fig6} does not contribute and furthermore there is only one way to extract a contribution from the diagram in Figure \ref{fig5}. Specifically, we take two powers of $\eta_2$ and  only one power of $\eta_1$ from the $\delta^{(4)}$ in \eqref{if}, while the remaining power of $\eta_1$ will come from differentiating the amplitude on the right-hand side of the recursion. 
Doing so we get 
\beqa
&&
{ \langle \hat{2} \hat{P} \rangle^3
\over 
\langle 1 \hat{2} \rangle \langle \hat{P} n\!+\!2 \rangle \langle n\!+\!2\, 1 \rangle}\, 
 \left({\langle 1 \hat{P} \rangle \over \langle \hat{2} \hat{P} \rangle} \right)
 \left({\langle n\!+\!2\, \,  \hat{P} \rangle \over \langle \hat{2} \hat{P} \rangle} \right)
 \left({\langle 1\,  \hat{2} \rangle \over \langle \hat{2} \hat{P} \rangle} \right) 
 \nonumber \\
 &&
\cdot \ \epsilon_{a_1 a_2 a_3 a_4}  \eta_2^{a_1}  \eta_2^{a_2} \eta_1^{a_3} \eta_{n\!+\!2}^{a_4}\, \eta_1^{a_5} {\partial \over \partial \eta_{\hat{P}}^{a_5}}  
\, 
A_{n} ( -\hat{P}, \hat{3}, \ldots n\!+\!1 )\, , 
\eeqa
which after using \eqref{relat} becomes simply
\beq
A_{n+2} \to {1\over p_{n+2} \cdot q_{12}}  \epsilon_{a_1 a_2 a_3 a_4} \eta_2^{a_1}  \eta_2^{a_2} \eta_1^{a_3} \eta_{n\!+\!2}^{a_4}\, \eta_1^{a_5} {\partial \over \partial \eta_{\hat{P}}^{a_5}}  
\, 
A_{n} ( -\hat{P}, g_3^{-} , \ldots n\!+\!1 )\,  , 
\eeq
 where we recall that we selected particle $3$ to be a gluon of negative helicity.  This contribution diverges as $1/\delta$ in the double-soft limit. We also note that this case is entirely similar to that discussed in \cite{ArkaniHamed:2008gz} (however note that in that case, particle $3$ was replaced by an auxiliary negative-helicity graviton, which was taken soft and decoupled at the end of the calculation).  

\vspace{1cm}

\subsubsection*{Acknowledgments}
We would like to thank  Lorenzo Bianchi, Massimo Bianchi, Andi Brandhuber, Ed Hughes, Bill Spence and   Congkao Wen for discussions on related topics.  GT thanks the  Institute for Physics, IRIS Adlershof 
and the Kolleg Mathematik und Physik at Humboldt University, Berlin,  as well as the Physics Department at the University of Rome ``Tor Vergata" for their warm hospitality and support. The work of GT  was supported by the Science and Technology Facilities Council Consolidated Grant ST/L000415/1  
{\it ``String theory, gauge theory \& duality".} The work of TMcL was supported in part by Marie Curie Grant CIG-333851.
DN's research is supported by the SFB 647 ``Raum-Zeit-Materie. Ana\-ly\-tische
und Geometrische Strukturen" grant.

\newpage

\appendix
\section{Sub-subleading terms}
\label{app:ss_csl}
We can continue our analysis of the double-soft terms in the gravitational case to the
sub-subleading terms. For the consecutive double-soft limit we have 
we have 
\<
{\tt CSL}^{(2)}(1^+, 2^\pm)=S^{(1)}(q_2^\pm)S^{(1)}(q_1^+)+S^{(0)}(q_2^\pm)S^{(2)}(q_1^+)+S^{(2)}(q_2^\pm)S^{(1)}(q_1^+)~.
\>
\paragraph*{The $1^+2^+$ case.}
A brief calculation shows that in the case of two positive helicity gluons 
\<
{\tt CSL}^{(2)}(1^+, 2^+)&=&-\frac{[12]}{\vev{12}^2}\sum_{a\neq 1, 2}\langle a|q_{12}|a]\frac{[2\partial_{\tilde a}]}{\vev{1a}}\nn\\
&&+\frac{1}{2\vev{12}^2}\sum_{a, b\neq 1,2}\frac{[2a][1b]}{\vev{2a}\vev{1b}}\big(\vev{1a}[1\partial_{\tilde b}]- \vev{2b}[2 \partial_{\tilde a}]\big)^2
\>
where we have used the notation $[1 \partial_{\tilde a}]={\tilde \lambda}_1^{\dot \alpha}\tfrac{\partial}{\partial {\tilde \lambda}_a^{\dot \alpha}}$ etc. 
Because of the contact term the antisymmetric combination is non-trivial and can be simplified to 
\<
{\tt aCSL}^{(2)}(1^+, 2^+)&=&-\frac{[12]}{2\vev{12}^2}\sum_{a\neq 1, 2}\left(\frac{\vev{1a}}{\vev{2a}}[1a][1\partial_{\tilde a}]-\frac{\vev{2a}}{\vev{1a}}[2a][2\partial_{\tilde a}]\right)~.
\>
\paragraph*{The $1^+2^-$ case.} 
For the mixed helicity case we find 
\<
{\tt CSL}^{(2)}(1^+, 2^-)&=&\frac{1}{[12]\vev{12}} \sum_{a\neq 1, 2} \frac{[1a]\vev{2a}^4}{\vev{1a}^3}\nn\\
&&+\sum_{a\neq 1,2} \frac{\vev{2a}^2[1a]}{[2a]\vev{1a}^2} \left(\frac{[1a]}{[12]} [1\partial_{\tilde a}]- \frac{\vev{2a}}{2\vev{21}}\vev{2 \partial_{ a}}\right)\nn\\
& &+\frac{1}{2}\sum_{a,b\neq 1,2} \frac{\vev{2a} [1b]}{[2a]\vev{1b}}\left(\frac{[1a]}{[12]}[1\partial_{\tilde b}]- \frac{\vev{2b}}{\vev{21}}\vev{2 \partial_{ a}}\right)^2
\>
where in the last line the expression should be understood with the derivatives always to the right, i.e. they don't act on the $\lambda/{\tilde \lambda}$'s
in the double-soft factor itself. Of particular interest is the first term which arises as a contact term but one where the derivatives act on the
soft momenta and so this term in fact has scaling behaviour of the same order as ${\tt CSL}^{(1)}$.

\section{Supersymmetric Yang-Mills soft limits}
\label{app:susy_soft}
It is straightforward to consider the supersymmetric generalisation of the previous calculations. Let us briefly review the single soft case in Yang-Mills. 
Given an $(n\!+\!1)$-point superamplitude the soft limit, with particle $1$ being soft, 
is naturally taken as 
\<
\{\lambda_1, { \tilde \lambda}_1, \eta_1\} \rightarrow \{ \sqrt{\delta} \lambda_1,   \sqrt{\delta} { \tilde \lambda}_1, \eta_1\}
\>
with $\delta \rightarrow 0$. In particular with this choice of scaling both $q=\sum_i \lambda_i \eta_i$ and $\tilde q=\sum_i {\tilde \lambda}_i \tfrac{\partial}{\partial \eta_i}$ 
scale identically.  Using the little transformation of the superamplitude, this implies 
\<
A_{n+1}( \{ \sqrt{\delta} \lambda_1,   \sqrt{\delta} { \tilde \lambda}_1, \eta_1\})= \delta A_{n+1}( \{ \delta \lambda_1,    { \tilde \lambda}_1, \tfrac{1}{\sqrt{\delta}}\eta_1\})~.
\>
However the analysis of this limit seems more complicated via BCFW due to the number of diagrams contributing. 
Instead we can consider, following \cite{He:2014bga,Nandan:2012rk}, 
\<
\label{eq:hol_susy}
\{\lambda_1, { \tilde \lambda}_1, \eta_1\} \rightarrow \{ \sqrt{\delta} \lambda_1,   \sqrt{\delta} { \tilde \lambda}_1, \sqrt{\delta} \eta_1\}~.
\>
Hence, after using the little scaling,  we find the holomorphic limit of the superamplitude,
\<
\lim_{\delta\to 0}A_{n+1}( \{ \delta \lambda_1,    { \tilde \lambda}_1, \eta_1\})&=&\Big[ \frac{1}{\delta^2} S^{(0)}(n, s, 2)+\frac{1}{\delta} S^{(1)}(n, s, 2) \Big]  A_n~\nn\\
&\equiv&S(n,s,2)A_n
\>
which defines the holomorphic soft factor $S(n,s,2)$  given by,  see \cite{He:2014bga}, 
\<
\label{eq:susy_soft}
S^{(k)}(n, s, 2)=\frac{1}{k!}\frac{\vev{n2}}{\vev{ns}\vev{s2}}\Big[\frac{\vev{sn}}{\vev{2n}}\left( {\tilde \lambda}_s\cdot \frac{\partial }{\partial{\tilde \lambda}_2}+ {\eta}_s\cdot \frac{\partial }{\partial{\eta}_2}\right)
+\frac{\vev{s2}}{\vev{ n2} }\left( {\tilde \lambda}_s\cdot \frac{\partial }{\partial{\tilde \lambda}_n}+ {\eta}_s\cdot \frac{\partial }{\partial{\eta}_n}\right)\Big]^k~.
\>
We can also consider the anti-holomorphic limit \cite{He:2014bga}, 
under which
 \<
\lim_{\delta\to 0}A_{n+1}( \{  \lambda_1,   \delta { \tilde \lambda}_1, \eta_1\})&=&\Big[ \frac{1}{\delta^2} {\bar S}^{(0)}(n, s, 2)+\frac{1}{\delta} {\bar S}^{(1)}(n, s, 2) \Big]  A_n~\nn\\
&\equiv&{\bar S}(n,s,2)A_n~, 
\>
where the anti-holomorphic soft factor is given by 
\<
{\bar S}^{(k)}(n,s,2)=\frac{1}{k!}\frac{[n2]}{[ns][s2]}\delta^{(4)}(\eta_s+\delta \frac{[ns]}{[2n]}\eta_2+\delta \frac{[s2]}{[2n]}\eta_n)\Big[\frac{[sn]}{[2n]}\lambda_s\cdot \frac{\partial}{\partial \lambda_2}+\frac{[s2]}{[n2]}\lambda_s\cdot \frac{\partial}{\partial \lambda_n}\Big]^k~.
\>


\bibliographystyle{nb}
\bibliography{soft}

\end{document}